\documentclass{aa}
\usepackage{multirow}
\usepackage{cancel}
\usepackage{comment}
\usepackage{amsmath}
\usepackage{yhmath}
\usepackage{amssymb}
\usepackage{bm}
\usepackage{graphicx}
\usepackage[caption=false]{subfig}
\usepackage{lscape}
\usepackage{longtable}
\usepackage{txfonts}
\usepackage{natbib}
\usepackage{booktabs}
\usepackage[dvipsnames,table,xcdraw]{xcolor}
\usepackage{xspace}

\makeatletter
\renewcommand*\aa@pageof{, page \thepage{} of \pageref*{LastPage}}

\newcommand{\erosita}{eROSITA\xspace}
\newcommand{\planck}{\emph{Planck}\xspace}

\newcommand{\esass}{\texttt{eSASS}\xspace}

\newcommand{\erass}[1][1]{eRASS1\xspace}

\newcommand{\Om}{\Omega_\textrm{m}}

\newcommand{\dd}{\mathrm{d}}
\newcommand{\logzc}{\log_{10} z_\textrm{c}}

\newcommand{\OmegameRASS}{0.28\pm0.02}
\newcommand{\sigmaEighteRASS}{0.89 \pm 0.03}
\newcommand{\SheRASS}{0.85\pm0.01}
\newcommand{\logzceRASS}{3.3^{+0.3}_{-0.7}}
\newcommand{\fEDEeRASS}{0.3}

\newcommand{\OmegameRASSACT}{0.32\pm 0.01}
\newcommand{\sigmaEighteRASSACT}{0.83\pm 0.01}
\newcommand{\SheRASSACT}{0.86 \pm 0.01}
\newcommand{\logzceRASSACT}{3.5^{+0.2}_{-0.3}}
\newcommand{\fEDEeRASSACT}{0.14}
\newcommand{\HZeroeRASSACT}{69.2^{+1.3}_{-2.4}}

\newcommand{\OmegameRASSPACT}{0.31 \pm 0.01}
\newcommand{\sigmaEighteRASSPACT}{0.84 \pm 0.01}
\newcommand{\SheRASSPACT}{0.85 \pm 0.01}
\newcommand{\logzceRASSPACT}{3.52 \pm 0.09}
\newcommand{\fEDEeRASSPACT}{0.09\pm0.04}
\newcommand{\HZeroeRASSPACT}{70.3^{+1.3}_{-1.5}}

\newcommand{\OmegameRASSPACTLB}{0.30 \pm 0.01}
\newcommand{\sigmaEighteRASSPACTLB}{0.84 \pm 0.01}
\newcommand{\SheRASSPACTLB}{0.84 \pm 0.01}
\newcommand{\logzceRASSPACTLB}{3.53 \pm 0.09}
\newcommand{\fEDEeRASSPACTLB}{0.10^{+0.04}_{-0.03}}
\newcommand{\HZeroeRASSPACTLB}{71.4 \pm 1.4}

\usepackage{twoopt}
\defcitealias{Ghirardini2024}{G24}
\defcitealias{PlanckCollaboration2020}{Planck20}
\defcitealias{Riess2022}{SH0ES}
\defcitealias{Riess2019}{SH0ES19}
\defcitealias{Louis2025}{ACT-DR6}
\defcitealias{Calabrese2025}{ACT-EDE}

\makeatletter
\newcommand{\lcdm}{\ensuremath{\Lambda\mathrm{CDM}}\xspace}
\newcommandtwoopt{\citeads}[3][][]{\href{http://adsabs.harvard.edu/abs/#3}%
    {\def\hyper@linkstart##1##2{}%
     \let\hyper@linkend\@empty\citealp[#1][#2]{#3}}}
  \newcommandtwoopt{\citepads}[3][][]{\href{http://adsabs.harvard.edu/abs/#3}%
    {\def\hyper@linkstart##1##2{}%
     \let\hyper@linkend\@empty\citep[#1][#2]{#3}}}
  \newcommandtwoopt{\citetads}[3][][]{\href{http://adsabs.harvard.edu/abs/#3}%
    {\def\hyper@linkstart##1##2{}%
     \let\hyper@linkend\@empty\citet[#1][#2]{#3}}}
  \newcommandtwoopt{\citeyearads}[3][][]%
    {\href{http://adsabs.harvard.edu/abs/#3}
    {\def\hyper@linkstart##1##2{}%
     \let\hyper@linkend\@empty\citeyear[#1][#2]{#3}}}
\makeatother

\usepackage{siunitx} 
\usepackage{hyperref}
\hypersetup{colorlinks=true,allcolors=[rgb]{0,0,0.8}}

\title{The SRG/eROSITA All-Sky Survey}
\subtitle{Early dark energy and Hubble constant from the cluster mass function }

\author{
  J.~Strunk\inst{1}
  \and E.~Artis\inst{1}
  \and E.~Bulbul\inst{1,2}
  \and S.~Grandis\inst{3}
  \and N.~Clerc\inst{4}
  \and V.~Ghirardini\inst{5}
  \and M.~Kluge\inst{1}
  \and A.~Liu\inst{6}
  \and F.~Balzer\inst{1}
  \and J. Comparat\inst{7}
  \and I.~Chiu\inst{8}
  \and Z.~Ding\inst{1}
  \and N.~Malavasi\inst{1}
  \and A.~Merloni\inst{1}
  \and T.~Mistele\inst{1}
  \and H.~Miyatake\inst{9,10,11}
  \and S.~Miyazaki\inst{12}
  \and K.~Nandra\inst{1}
  \and F.~Kleinebreil\inst{3}
  \and N.~Okabe\inst{13}
  \and M.~E.~Ramos-Ceja\inst{1}
  \and J.~S.~Sanders\inst{1}
  \and T.~Schrabback\inst{3}
  \and R.~Seppi\inst{14}
  \and S.~Zelmer\inst{1}
  \and X.~Zhang\inst{1}
}

\institute{
  Max Planck Institute for Extraterrestrial Physics, Giessenbachstrasse 1, 85748 Garching, Germany \label{inst1}
  \and
  Fakult\"at f\"ur Physik, LMU M\"unchen, Schellingstr. 4, 80799 München, Germany \label{inst2}
  \and
  Universit\"at Innsbruck, Institut f\"ur Astro- und Teilchenphysik, Technikerstr. 25/8, 6020 Innsbruck, Austria \label{inst3}
  \and
  IRAP, Université de Toulouse, CNRS, UPS, CNES, F-31028 Toulouse, France \label{inst4}
  \and
  INAF, Osservatorio di Astrofisica e Scienza dello Spazio, via Piero Gobetti 93/3, I-40129 Bologna, Italy \label{inst5}
  \and
  Institute for Frontiers in Astronomy and Astrophysics, Beijing Normal University, Beijing 102206, China\label{inst6}
  \and
  Université Grenoble Alpes, CNRS, Grenoble INP, LPSC-IN2P3, 53, Avenue des Martyrs, 38000, Grenoble, France\label{inst7}
  \and
  Department of Physics, National Cheng Kung University, 70101 Tainan, Taiwan \label{inst8}
  \and
  Kobayashi-Maskawa Institute for the Origin of Particles and the Universe (KMI), Nagoya University, Nagoya, 464-8602, Japan \label{inst9}
  \and
  Institute for Advanced Research, Nagoya University, Nagoya 464-8601, Japan \label{inst10}
  \and
  Kavli Institute for the Physics and Mathematics of the Universe (WPI), The University of Tokyo Institute for Advanced Study (UTIAS), The University of Tokyo, Chiba 277-8583, Japan \label{inst11}
  \and
  Subaru Telescope, National Astronomical Observatory of Japan, 650 N Aohoku Place Hilo, HI 96720 USA \label{inst12}
  \and
  Department of Physical Science, Hiroshima University, 1-3-1 Kagamiyama, Higashi-Hiroshima, Hiroshima 739-8526, Japan \label{inst13}
  \and
  Department of Astronomy, University of Geneva, Ch. d’Ecogia 16, CH-1290 Versoix, Switzerland \label{inst14}
}
\date{\today}

\titlerunning{SRG/eROSITA early dark energy and Hubble constant from the cluster mass function}
\authorrunning{Strunk et al.}

\begin{document}

\abstract{
The evolution of the mass function of massive galaxy clusters is a well-established probe of cosmology. Using the eROSITA X-ray instrument, on board the \textit{Spectrum Roentgen Gamma} (SRG) mission, the western Galactic hemisphere of eROSITA’s first All-Sky Survey (eRASS1) delivers a uniformly selected and securely confirmed sample of 5259 galaxy clusters spanning $0.1 < z < 0.8$. 
When combined with overlapping weak lensing data from DES Year 3, KiDS, and HSC for mass calibration, this dataset enables precise tests of the standard \lcdm framework and beyond. In this work, we use the $0.1 < z < 0.45$ subsample of the eRASS1 galaxy cluster catalog to place constraints on an axion-like Early Dark Energy (EDE) scenario. Such models have been proposed as a possible mechanism to alleviate the tension between early- and late-Universe measurements of the Hubble constant. In this framework, a scalar field temporarily enhances the cosmic expansion rate around the epoch of recombination, then rapidly dilutes at later times, leaving distinct imprints on the growth of structure. We leverage these signatures across a redshift and scale range that has been scarcely studied in the literature to present the first constraints on EDE derived from galaxy cluster number counts. Using eRASS1 number counts alone, we obtain an upper limit on the maximum EDE fraction of $f_\mathrm{EDE} < 0.3$, consistent with results from primary CMB analyses. This demonstrates that the cluster mass function provides a powerful and complementary probe of the EDE parameter space. Combining our cluster analysis with primary CMB measurements, baryon acoustic oscillations, and CMB lensing, while notably excluding distance-ladder calibration data, yields results consistent with a nonzero EDE contribution and $H_\mathrm{0}$ values compatible with late-Universe measurements. The most significant detection, at $\sim3.2\sigma$, arises from the joint analysis of the eRASS1 cluster sample with the full set of external datasets. This yields $f_\mathrm{EDE} = \fEDEeRASSPACTLB$, places competitive constraints on the critical redshift parameter, $\logzc = \logzceRASSPACTLB$, and leads to an inferred value of the Hubble constant $H_\mathrm{0}=\HZeroeRASSPACTLB \;{\rm km}\,{\rm s}^{-1}\,{\rm Mpc}^{-1}$. }

\keywords{(galaxies:) clusters: general --
(galaxies:) clusters: intracluster medium --
(cosmology:) cosmological parameters --
(cosmology:) observations --
(cosmology:) dark matter --
(cosmology:) large-scale structure of the Universe --}

\maketitle
\section{Introduction}
\label{sec:intro}

Within the standard flat $\Lambda$ Cold Dark Matter paradigm  (\lcdm), recent improvements in the precision of cosmological parameter measurements have revealed a significant discrepancy between direct and indirect determinations of the expansion rate, i.e., the Hubble constant, $H_\mathrm{0}$ \citep[e.g.,][]{Knox2020, DiValentino2021}. 
Direct measurements of the expansion rate in the late Universe rely on calibrating the intrinsic properties of astrophysical sources or utilizing independent geometric probes to determine cosmic distances. Currently, the most precise constraints are obtained using standard candles, typically Type Ia Supernovae (SN~Ia) calibrated with Cepheid variables \citep[e.g.,][ \citetalias{Riess2022} hereafter]{Riess2022}, or the Tip of the Red Giant Branch \citep[e.g.,][]{Scolnic2023, Freedman2025}. Complementarily, indirect measurements infer $H_\mathrm{0}$ by leveraging the model-dependent early-Universe sound horizon scale as a standard ruler. For example, this can be done using data from the Cosmic Microwave Background \citep[CMB; e.g.,][]{PlanckCollaboration2020d, Louis2025, Camphuis2025}, or from the imprint of Baryonic Acoustic Oscillations (BAO) on the Large-Scale Structures (LSS), combined with Big Bang Nucleosynthesis (BBN) constraints \citep[e.g.,][]{Schoneberg2022}. The statistical significance of the discrepancy between the two measurement methods ranges from $\sim 4-6 \sigma$, depending on the specific probes and combinations being compared \citep{Verde2019}.

Although physical explanations exist, this tension may also result from systematic effects in the analyses of the different individual datasets. These have been extensively investigated and reviewed in the literature \citep[e.g.,][]{Shah2021, DAmico2021, Kenworthy2022, Riess2022, Riess2024, Freedman2023}, yet the significant trend towards higher $H_\mathrm{0}$ observed in direct measurements seemingly remains unchanged after a careful examination of systematic uncertainties \citep{Verde2024}. This suggests that the discrepancy either originates from unknown systematics or from the need to extend \lcdm by considering new physical models. To be viable, these extensions must resolve the expansion rate discrepancy while also replicating \lcdm's success in describing other cosmological observations. \cite{Efstathiou2021} showed that the late-time observations strongly constrain changes in low redshift fundamental physics, with departures unlikely to be able to resolve the tension. Consequently, as highlighted in \citet{Knox2020}, a viable and relatively less constrained alternative could come from new physics in the early Universe ($z\gtrsim 1100$) that increases the model-dependent inferred $H_\mathrm{0}$ value. 
 
 The axion-like Early Dark Energy (EDE) scenario provides a potential solution to the $H_\mathrm{0}$ tension by introducing a scalar field component that temporarily contributes to the energy budget of the Universe near matter-radiation equality, before dissipating with expansion faster than matter \citep{Poulin2018, Smith2020}. By briefly increasing the pre-recombination expansion rate, EDE naturally reduces the sound horizon at decoupling $r_\mathrm{s}(z_*)$ (Equation~\ref{eq:r_s}), thereby increasing the model-dependent value of $H_\mathrm{0}$ inferred from early-Universe data. This modification leaves the late-time expansion history relatively unchanged, providing a phenomenologically motivated framework to alleviate the tension while maintaining consistency with \lcdm constraints from low-redshift probes. 
 
 The EDE model is often implemented through the following parameters: the initial value of the field, $\theta_\mathrm{i}$; the maximum fractional energy density contribution to the expansion rate, $f_\mathrm{EDE}$; and the critical redshift where the maximum contribution is reached, $z_\mathrm{c}$. At the background level, EDE increases the expansion rate before recombination, thereby increasing Hubble friction and thus suppressing structure growth on sub-horizon scales while the EDE-induced boost is still relevant. The overall strength of the suppression is primarily given by $f_\mathrm{EDE}$. The parameter $\logzc$ sets the relevant horizon scale, whereas $\theta_i$ largely determines the EDE perturbation dynamics, imprinted on cosmological observables for modes that enter the horizon around $\logzc$ \citep{Poulin2023}. 

 To maintain the excellent fit to the CMB data, the EDE-induced scale-dependent suppression is compensated by increases in the physical present-day cold dark matter (CDM) density $\omega_\mathrm{CDM}=\Omega_\mathrm{CDM}h^2$, the spectral index $n_\mathrm{s}$, and the scalar amplitude $A_\mathrm{s}$ \citep{Smith2020, Ivanov2020}. Furthermore, since EDE cosmologies behave like \lcdm at late times, the constraints on $\Omega_\mathrm{CDM}$ imposed by BAO and SN~Ia data also induce a shift in $\omega_\mathrm{CDM}$. Given the relation between $\Omega_\mathrm{CDM}$ and $h^2$, the increase in $h^2$ in EDE cosmologies leads to an increase in $\omega_\mathrm{CDM}$, independently of the impact of EDE on structure growth. In this sense, the suppression of growth by EDE provides a mechanism to balance the enhanced growth due to the increase in $\omega_\mathrm{CDM}$ imposed by the cosmic calibration tension \citep{Poulin2024}. 

\begin{figure}[htb]
\centering
    \includegraphics[width=\linewidth]{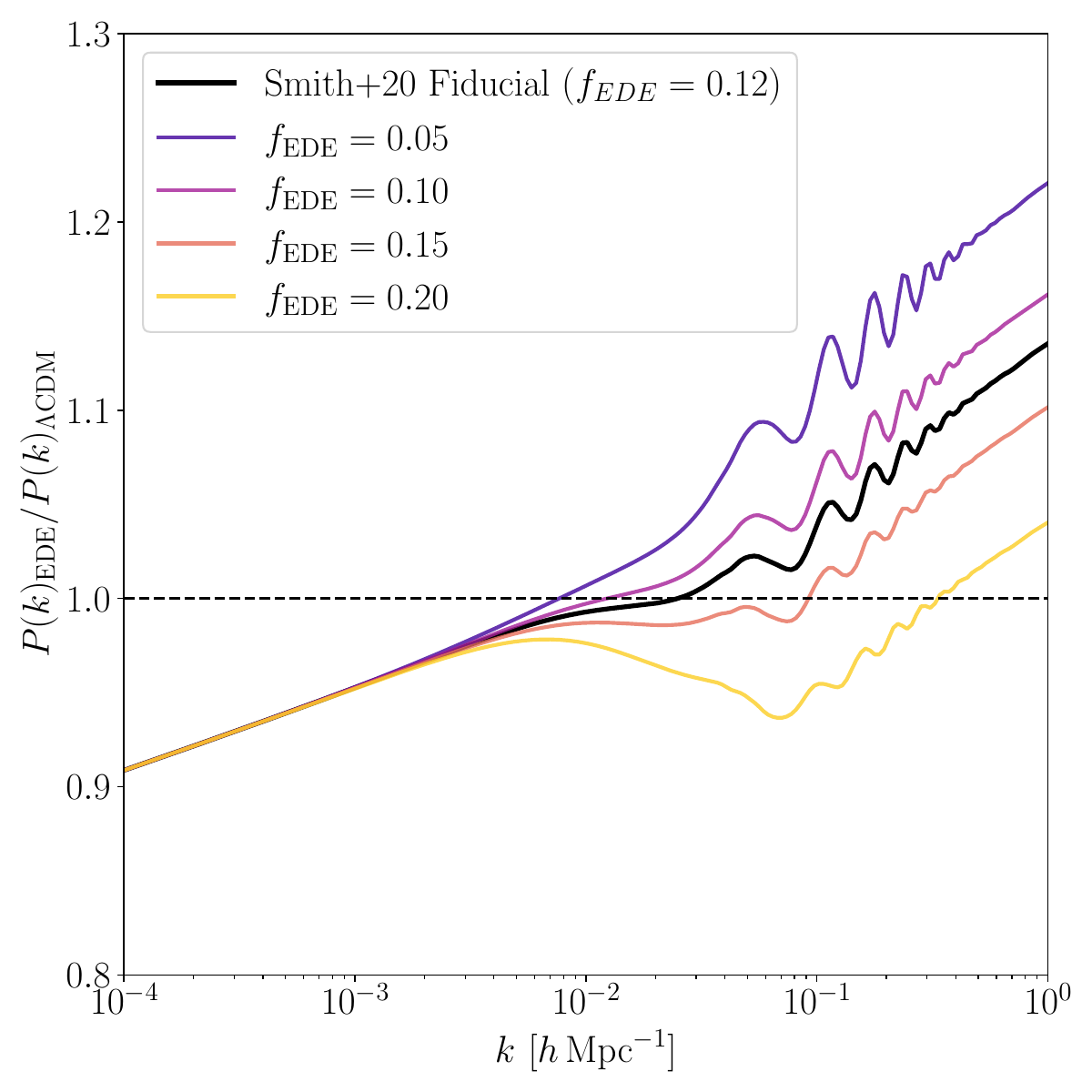}
    \caption{Relative difference in the linear matter power spectrum between EDE cosmologies and \lcdm (based on the \planck best-fit parameters). The fiducial EDE cosmology, plotted in black, is evaluated at the best-fit parameters reported in \citet{Smith2020}. Additional fractions of early dark energy are also plotted, showing enhanced small-scale power suppression with increasing $f_\mathrm{EDE}$.  }
    \label{fig:Fig1}
\end{figure}

The resulting EDE matter power spectrum is a balance between the unique suppression of growth of EDE and the enhancement induced by increases in $n_s$, $A_s$, and $\omega_\mathrm{CDM}$. 
Due to these shifts, as can be seen in black in Fig.~\ref{fig:Fig1} in relation to \lcdm, the EDE linear matter power spectrum features more power on small scales $k\geq k_\mathrm{c}$, where $k_\mathrm{c}$ is the wavenumber corresponding to the horizon scale at $z_\mathrm{c}$. With an increase in $f_\mathrm{EDE}$ relative to the fiducial EDE cosmology, the EDE-induced suppression at small scales is amplified, opposing the excess induced by the parameter shifts relative to \lcdm. 
 
Analyses utilizing \citetalias{Riess2022}-calibrated SN~Ia data have shown that a peak EDE energy budget of $f_\mathrm{EDE}\sim0.1$ can increase the inferred $H_\mathrm{0}$ constraints to be compatible with late-time geometric probes \citep{Smith2020}, however, primary CMB data alone do not show a strong preference for EDE \citep{Hill2020, Efstathiou2023, Khalife2025}. Considering a combination of the Atacama Cosmology Telescope (ACT) Data Release 4 observations \citep{Choi2020}, \planck data \citep[hereafter \citetalias{PlanckCollaboration2020}]{PlanckCollaboration2020} and BAO information from multiple surveys \citet{Hill2022} found  $f_\mathrm{EDE}=0.091^{+0.020}_{-0.036}$, and $H_\mathrm{0} = 70.9^{+1.0}_{-2.0}\;{\rm km}\,{\rm s}^{-1}\,{\rm Mpc}^{-1}$. However more recently, the use of ACT Data Release 6 \citep[hereafter \citetalias{Louis2025}]{Louis2025} with a \citetalias{PlanckCollaboration2020} prior on the optical depth, $\tau_\mathrm{reio}$, strongly relaxed this apparent preference, finding instead $f_\mathrm{EDE}<0.09$, and $H_\mathrm{0} = 67.9^{+0.9}_{-1.7}\;{\rm km}\,{\rm s}^{-1}\,{\rm Mpc}^{-1}$ \citep[hereafter \citetalias{Calabrese2025}]{Calabrese2025}. Combined with \citetalias{PlanckCollaboration2020} temperature and polarization data, the constraints are $f_\mathrm{EDE}<0.12$, and $H_\mathrm{0} = 69.3^{+0.9}_{-1.5}\;{\rm km}\,{\rm s}^{-1}\,{\rm Mpc}^{-1}$.

A consensus on the constraints on EDE when CMB data are jointly analyzed with additional probes has yet to emerge, underscored by the discrepancy observed when \citetalias{Riess2022}-calibrated SN~Ia data are included.
In Bayesian analyses, using the \planck \texttt{NPIPE} pipeline, which produces frequency maps from time ordered measurements for the CMB temperature and polarization data \citep{PlanckCollaboration2020c}, together with BAO, uncalibrated SN~Ia and Redshift Space Distortions (RSD), \cite{Efstathiou2023}, found an upper limit of $f_\mathrm{EDE}<0.061$, together with $H_\mathrm{0} = 68.11^{+0.47}_{-0.82}\;{\rm km}\,{\rm s}^{-1}\,{\rm Mpc}^{-1}$. In contrast, the inclusion of the \citetalias{Riess2022}-calibration in the SN~Ia measurements in their Bayesian analyses leads to $f_\mathrm{EDE}=0.107\pm0.023$. This approach is common in the EDE literature when including SN~Ia data and leads to non-zero constraints on $f_\mathrm{EDE}$ \citep[e.g.][]{Smith2020, Gsponer2024, Poulin2025}. However, the implicit inclusion of the late-time $H_\mathrm{0}$ measurement cannot test concordance, i.e., whether an EDE analysis independent of direct geometric probes is able to predict a higher $H_\mathrm{0}$ consistent with late-time constraints \citep{Hill2020}.

Because LSS probes feature parameter degeneracies orthogonal to those of the CMB, they offer an independent and complementary test of EDE via its impact on structure growth. The net effect of EDE-induced suppression and the excess brought about by the parameter shifts required to fit the CMB can be broadly summarized through an increase in $\sigma_\mathrm{8}$, the root mean square of fluctuations at $z=0$ filtered at a scale $R=8 h^{-1}\mathrm{Mpc}$ \citep{McDonough2024}. Observationally, this directly impacts the parameter $S_8\equiv\sigma_8\sqrt{\Om/0.3}$, the approximate degeneracy direction along which LSS probes jointly constrain $\Om$ and $\sigma_8$. In EDE cosmologies studied with CMB data, the $S_8$ value is found to be larger than in equivalent \lcdm analyses, and positively correlated with $f_\mathrm{EDE}$. 
Given that cosmic shear and galaxy clustering probes routinely have found $2-3\sigma$ lower values of $S_8$ than inferred from CMB experiments within \lcdm \citep[known as the $S_8$ discrepancy, see e.g.][for detailed reviews]{Abdalla2022,Perivolaropoulos2022,DiValentino2025}, joint CMB and LSS analyses have frequently argued that LSS data disfavor EDE as a solution to the Hubble tension \citep[e.g.,][for a review of the inclusion of LSS data in EDE analyses]{McDonough2024}.  

For a flat EDE cosmology, \cite{Hill2020} showed that the combination of primary CMB, CMB Lensing, BAO, RSD, and SN~Ia together with weak lensing (WL) shear data leads to a stringent constraint of $f_\mathrm{EDE}<0.06$.  
\cite{Ivanov2020} report constraints using the full-shape spectrum likelihood of the Baryon Oscillation Spectroscopic Survey \citep[BOSS,][]{Alam2017}, through the Effective Field Theory of LSS \citep[EFTofLSS][]{Baumann2012, Carrasco2012}, to obtain $f_\mathrm{EDE}<0.072$. In a similar manner, \cite{Gsponer2024} combined eBOSS and BOSS data with \planck, external BAO, and Pantheon+ SN~Ia data to obtain $f_\mathrm{EDE}<0.0584$. \citet{Qu2024b} used the angular power spectrum of \citetalias{PlanckCollaboration2020}, CMB lensing from \citet{PlanckCollaboration2020b} and \cite{Louis2025} and BAO from the Dark Energy Spectroscopic Instrument \citep[DESI,][]{Adame2025} to constrain $f_\mathrm{EDE}<0.09$. The inclusion of LSS data in Bayesian analyses, therefore, suggests that the EDE energy budget required to significantly reduce the Hubble tension is incompatible with the structure growth parameters constrained by late-Universe observables. 

Yet the conclusions drawn from the inclusion of LSS data might not be as straightforward, with re-analyses of BOSS data relaxing the constraints on $f_\mathrm{EDE}$ \citep{Simon2023} or showing dependence on the choice of priors on the EFTofLSS parameters \citep{Holm2023}. Moreover, Bayesian analyses that exclude the \citetalias{Riess2022} measurements have been argued to be prone to prior volume effects, leading to constraints on $f_\mathrm{EDE}$ that potentially mask the model's ability to resolve the Hubble tension \citep[e.g.,][]{Smith2021, Herold2022}. Alternatively, profile likelihood analyses circumvent this potential bias and lead to results relaxing the CMB constraints obtained on $f_\mathrm{EDE}$  \citep{Herold2022,Efstathiou2023} or even favoring a non-zero EDE component at $>2\sigma$ confidence for different combinations of multiprobe datasets \citep{Herold2023, Poulin2025}. 
Consequently, the constraints on EDE and its impact on the $H_\mathrm{0}$ tension remain debated. 

In this work, we present the first constraints on EDE obtained using 4196 clusters of galaxies detected in the first eROSITA all-sky survey (eRASS1) \citep{Bulbul2024, Kluge2024}, in the redshift range $0.1 < z < 0.45$. The extragalactic area used in this work covers an area of 12791~deg$^{2}$ in the Western Galactic Hemisphere of the eRASS1 survey observed between December 12, 2019, and June 11, 2020 \citep{Predehl2021, Merloni2024}. The weak lensing mass calibration method performed with the overlapping surveys DES~Y3 \citep{Grandis2024a}, KiDS \citep{Kleinebreil2024}, and HSC Survey \citep{Okabe2025, Chiu2025}, combined with well-understood selection function modeling based on mock-observations \citep{Comparat2019, Seppi2022, Clerc2024}, enables robust constraints on EDE models from cluster number counts. 
The EDE halo mass function model is derived from the Press-Schechter formalism \cite{PressSchechter1974} using an EDE-modified linear matter power spectrum, and is integrated into our cosmology pipeline \citep{Ghirardini2024}. The models used in this work are verified against the set of simulations available in the literature \citep{Klypin2021}. 

This paper is organized as follows. In Section \ref{sec:theory} we describe the axion-like EDE model and its impact on structure formation. The datasets used are described in Section~\ref{sec:erass1}, and Section~\ref{sec:methods} gives an overview of the inference pipeline. Our results are presented in Section~\ref{sec:results}. Reported uncertainties correspond to a 68\% confidence level. The quoted upper limits are at a 95\% confidence level unless otherwise stated.

\section{Early dark energy cosmology}
\label{sec:theory}
This section introduces the general mechanism by which EDE models seek to resolve the Hubble tension and provides a phenomenological description of the axion-like model. 

\subsection{Impact on the sound horizon}

EDE models alleviate the Hubble tension by increasing the inferred value of $H_\mathrm{0}$ from early-Universe probes while keeping the late-time expansion history largely unchanged, e.g., as constrained by SN~Ia and BAO data at lower redshifts. This is achieved through a boost of the expansion rate $H(z)$, described through:

\begin{equation}
\label{eq:H_def}
H^2(z) = H_\mathrm{0}^2  \left[
    \Omega_\mathrm{m}(1+z)^3
    + \Omega_\mathrm{r}(1+z)^4
    + \Omega_{\Lambda}
    + \Omega_{\mathrm{EDE}}(z)
\right],
\end{equation}
with the fractional EDE density $\Omega_\mathrm{EDE}(z)$ describing the energy contribution of EDE to the background evolution, and the present-day density parameters $\Omega_{i}$, $\Omega_\mathrm{m}$ for matter, $\Omega_\mathrm{r}$ for radiation, and the cosmological constant dark energy ($\Lambda$), respectively. If the EDE contribution is localized in redshift before recombination, with $\Omega_\mathrm{EDE}(z\lesssim z_*)\to0$\footnote{$z_*$ marks the redshift during recombination at which photons decouple from matter}, decaying rapidly afterward, the physical size of the sound horizon, $r_\mathrm{s}(z)$, at decoupling $z_*$, 
\begin{equation}
\label{eq:r_s}
    r_\mathrm{s}(z_*)=\int_{z_*}^\infty{c_\mathrm{s}(z')}/{H(z')}\;\dd z',
\end{equation} 

is reduced without a significant modification of the late-time expansion $H(z\lesssim z_*)$. Given that the angular size of the sound horizon at decoupling, equivalent to the ratio of $r_{s}(z_*)$ and the angular diameter distance to the last scattering surface $D_A(z_*)$:

\begin{equation}
\theta_\mathrm{s}(z_*)=\frac{r_\mathrm{s}(z_*
)}{D_A(z_*)},
\end{equation} 
is strongly constrained by CMB data, the reduction in $r_\mathrm{s}(z_*)$ also brings a reduction to $D_\mathrm{A}(z_*)\propto \int^{z_*}_0dz/H(z)$. 
As such, for a flat EDE cosmology, $D_\mathrm{A}(z_*)$ depends through $H(z)$ (Equation~\ref{eq:H_def}) on $H_\mathrm{0}$ as well as the present day density parameters $\Omega_{i,0}$, which can be constrained independently of the sound horizon scale to high-precision by CMB data. An increase in $H(z\gtrsim z_*)$ through a non-negligible EDE contribution $\Omega_\mathrm{EDE}(z\approx z_\mathrm{c})\sim0.1$ leads thereby to an increase in the CMB-inferred $H_\mathrm{0}$ value, alleviating the tension with the higher late-time measurements.

In order to explain a component of the Universe capable of producing such a signature on the expansion history, EDE models rely on the existence of a certain scalar field $\phi$, minimally coupled to the metric $g^{\mu\nu}$. One can compute the background energy-momentum tensor of the scalar field and compare it to the energy-momentum tensor of a perfect fluid, and obtain the following energy density and pressure of the scalar field \citep{Poulin2018}:

\begin{equation}
    \rho_\phi=\frac{\dot{\phi}^2}{2}+V(\phi), \; p_\phi=\frac{\dot{\phi}^2}{2}-V(\phi) .
\end{equation}

The evolution of the field is governed by the homogeneous Klein-Gordon (KG) equation:

\begin{equation}
\label{eq:klein_gordon}
\ddot{\phi} + 3H\dot{\phi} + \dfrac{\dd V}{\dd \phi} = 0 \; .
\end{equation}
\subsection{Axion-like early dark energy}

For this work, we restrict ourselves to the axion-like EDE, with a potential of the form:
\begin{equation}
    V(\phi)=m^2f^2\left[1-\cos(\phi/f)\right]^n ,
\label{eq:pot}
\end{equation}
where $m$ is the axion mass and $f$ the axion decay constant. 

The peak contribution of the EDE field to the background expansion, at the critical redshift $\logzc$, is used to define the phenomenological parameter $f_\mathrm{EDE}\equiv\Omega_\mathrm{EDE}(z_\mathrm{c})$. Furthermore, the renormalized field variable $\theta\equiv\phi/f$ is defined to describe the misalignment angle such that $0\leq\theta\leq\pi$ without loss of generality. This EDE scenario essentially represents a flexible toy potential with several possible theoretical explanations \citep{Poulin2023}. These include higher-order instanton corrections \citep{Abe2015,Choi2016,Kappl2016}, axion-dilaton interactions \citep{Alexander2019} or explanations within the context of string theory \citep{Nakagawa2023,Cicoli2023}. This flexibility, combined with its relatively simple phenomenological parametrization, has made it a popular EDE model in the literature. 

At early times, the expansion rate $H(z)$ is large, so $H \gg m$. The Hubble friction term $3H\dot{\phi}$ in Equation~\ref{eq:klein_gordon} dominates over the potential gradient term, $\dd V(\phi)/\dd\phi$, leading to $\ddot{\phi}+3H\dot{\phi}\approx 0$. During this phase, where $z\gg z_\mathrm{c}$, the EDE field is frozen with $\dot{\phi}\approx0$, and contributes to the expansion rate similarly to a cosmological constant with equation of state parameter $w_\mathrm{EDE}$:
\begin{equation}
    w_\mathrm{EDE}=\frac{p}{\rho}=\frac{\frac{\dot{\phi}^2}{2}-V(\phi)}{\frac{\dot{\phi}^2}{2}+V(\phi)}\approx-1.
\end{equation}

  Given that the Hubble parameter decreases with expansion during the radiation- and matter-dominated eras, the potential-gradient term becomes relevant around $z_\mathrm{c}$, making the field dynamical. After, the KG equation (Equation~\ref{eq:klein_gordon}) becomes $\ddot{\phi}+\frac{\dd V}{\dd \phi}\approx0$, and the field oscillates. For potentials of the form $V(\phi)\sim\phi^{n}$, the effective equation of state parameter, which determines the time-averaged rate at which the field dissipates, is determined by $n$ through \citep{Poulin2018}: 
\begin{equation}
\label{eq:w_f}
    \langle w_\mathrm{EDE}\rangle=\frac{n-1}{n+1},
\end{equation}
where $n$ is the exponent of the potential in Equation~\ref{eq:pot}.

After $z_\mathrm{c}$, the EDE background energy density dilutes as rapidly as matter for $n=1$, like radiation for $n=2$, and faster than both for $n\geq 3$. Given that the EDE field must dissipate quicker than matter to leave the low-redshift expansion history unchanged, we require $n\geq 2$, thus excluding the standard axion with $n=1$, studied for eRASS1 data in the ultra light regime in \citet{Zelmer2025}. 

The initial field displacement $\theta_\mathrm{i}$ effectively controls the oscillation frequency of the background field. Because of this, it determines the effective sound speed $c_\mathrm{s}^2$ of the EDE perturbation and, to a great extent, its dynamical evolution \citep{Poulin2019}.
Described up to linear order as $\phi=\bar{\phi}+\delta\phi$, and once again describing the perturbations according to the continuity and Euler equations, the KG equation (Equation~\ref{eq:klein_gordon}) in synchronous gauge and Fourier space is \citep{Smith2020}:

\begin{equation}
    \delta\ddot{\phi}+3H\delta\dot{\phi}+\left(\frac{k^2}{a^2}+\frac{d^2V}{d\phi^2}\right)\delta\phi(k)=-\frac{\dot{\psi}\dot{\bar{\phi}}}{2} ,
\end{equation}
\noindent where $\psi$ is the trace of the spatial-spatial metric perturbation. EDE perturbations on subhorizon scales thus evolve as a driven damped harmonic
oscillator with an effective sound speed $c_\mathrm{s}$ \citep{Smith2020}. Consequently, $c_\mathrm{s}$, which can be related to the effective angular frequency $\omega_\mathrm{eff}=\sqrt{k^2+a^2\dd^2 V/\dd\phi^2}$ and is determined by $\theta_\mathrm{i}$, thus controls the balance between modes entering the horizon after, close to, and before $z_\mathrm{c}$ \citep{Poulin2023}. Given the minimal coupling to the metric, the effects of the EDE perturbations that will be felt by standard model particles are limited to modifications to the Weyl potential \citep{Poulin2023}. These equations are solved using standard Boltzmann solvers modified to include EDE. For this work, we use \texttt{CAMB} \texttt{EarlyQuintessence} \citep{Smith2020}, an EDE modification of the original \texttt{CAMB} code \citep{Lewis2000}. Other implementations include \texttt{CLASS\_EDE} \citep{Hill2020} and \texttt{AxiCLASS} \citep{Poulin2019, Smith2020}, both of which are built upon the \texttt{CLASS} solver \citep{Blas2011}. 

The EDE field dynamics suppress the growth of density perturbations, altering the matter power spectrum on sub-horizon scales before the EDE component dissipates. Consequently, the predicted abundance of dark matter halos deviates from \lcdm. Our theoretical model for the number density of DM halos, the Halo Mass Function (HMF), in an EDE cosmology is presented in Sec.~\ref{sec:hmf}, along with a discussion of its predictions for CMB-fitted EDE cosmologies.

\section{Survey data}
\label{sec:erass1}

This section briefly summarizes the X-ray and WL data used to constrain the EDE model parameters in this work. We direct the reader to the papers referenced below for a more detailed discussion of the analysis of the X-ray and WL data.

\subsection{\erosita cluster catalogs}
\label{subsec:erosita}

In this work, we utilize the volume-limited cosmology sample of 4196 clusters of galaxies detected in the first eROSITA all-sky survey, eRASS1, in the redshift range $0.1 < z < 0.45$ with a reported purity of 94.3\% \citep{Bulbul2024}. We note that the left-over contamination is modeled in the analysis and described in detail in Section~\ref{sec:mixtureModel} and in \citet{Kluge2024}. The cosmology sample, the largest ICM-selected cluster sample utilized for cosmological analyses to date, covers a sky area of 12,791~deg$^2$. The optical photometry and confirmation are based on data from the common footprint of eROSITA and the DESI Legacy Survey DR10-South Surveys in the Western Galactic Hemisphere \citep{Kluge2024}. 
The soft X-ray count rate, $C_\mathrm{R}$, is measured in the 0.2-2.3~keV band by the forward modeling imaging analysis tool MultiBand Projector 2D\footnote{https://github.com/jeremysanders/mbproj2d} \citep[MBProj2D;][]{Sanders2018, Liu2022}. The eROSITA X-ray data is reprocessed with the \texttt{eSASS} software as described in \citet{Merloni2024} by applying a correction for the Galactic absorption and a more accurate ICM and
background modeling. The details of the X-ray processing are provided in \citet{Bulbul2024, Liu2022}. The count-rate measurements, proportional to the integrated surface brightness, scale with halo mass and redshift. Consequently, this observable is used as a mass proxy in this work, following the method adopted by Ghirardini et al. (\citeyear{Ghirardini2024}, hereafter G24) and \citet{Artis2024, Artis2025,Zelmer2025}.

Based on findings reported in \citet{Artis2025}, we restrict our analysis to the local galaxy cluster eRASS1 sample ($0.1<z<0.45$) rather than the full cosmological sample ($0.1<z<0.8$). The highest redshift bin $(0.45 < z < 0.8)$ slightly shifts the values of $\sigma_8$ and $S_8$ towards higher values, and increases the scatter in the X-ray count-rate scaling relations. This effect is attributed to a higher fraction of contaminants at $z>0.45$, potentially driven by an increased AGN fraction in the sample \citep{Kluge2024}. To enable a consistent combination of our results with CMB probes, agreement of the cosmological parameters at the $2\sigma$ level is required. We therefore adopt the local galaxy cluster sample within 
$0.1<z<0.45$, where full consistency with $\Lambda$CDM is satisfied. 

\subsection{Weak lensing survey data}
\label{subsec:weak_lensing_survey}

To infer the halo mass function from cluster number counts, a key ingredient is the cluster mass, which is not a direct observable in X-ray surveys. Instead, such surveys rely on well-calibrated scaling relations between observable quantities and cluster mass to construct the halo mass function and model its redshift evolution. The calibration of the cluster scaling relations is achieved by incorporating WL measurements from deep, wide-area surveys with overlapping footprints with eROSITA: DES \citep{Gatti2021, Sheldon2017, Huff2017}, KiDS \citep{Kuijken2019, Wright2020, Hildebrandt2021, Giblin2021}, and the HSC Survey \citep{Aihara2018}. The tangential shear profiles of background galaxies around detected eRASS1 galaxy clusters serve as reliable mass indicators and are then used in the scaling relations.

The overlapping sky area between eROSITA and DES is 4,060 deg$^2$. The DES Y3 shape catalog \citep{Gatti2021}, built from observations in the $r,\;i,\,z$-bands, was used to produce the tangential shear profile for the 2201 eRASS1 galaxy cluster with a signal-to-noise ratio (SNR) of 65, presented in \cite{Grandis2024a}. The HSC survey in the $g,\;r,\;i,\;z,\;Y$ bands \citep{Aihara2018}, has a common area of $\approx500$ deg$^2$ with eROSITA. The HSC Y3 shape catalog \citep{Li2022} was used to obtain shear profiles, lensing covariance matrices, and redshift distributions for 96 eRASS1 galaxy clusters with a SNR of 40 \citep{Okabe2025, Chiu2025}. On the other hand, the KiDS survey, with observations in the $u,\;g,\;r,\;i$ bands, is used for WL and photometric redshift measurements. Its fourth data release, hereinafter KiDS-1000 \citep{Kuijken2019, Giblin2021, Hildebrandt2021, Wright2020}, is used to measure tangential profiles for 236 eRASS1 clusters (101 clusters in the KiDS-North field and 136 in the KiDS-South field) with a SNR of 19 \citep[for details, see][]{Kleinebreil2024}.

\section{Cosmology inference and implementation of the EDE models}
\label{sec:methods}

The cosmology pipeline utilized in eRASS1 is designed to simultaneously constrain the cosmological parameters, the scaling relation parameters required to calibrate the cluster mass scale, 
and the contamination modeling parameters \citepalias{Ghirardini2024}. It accounts for several aspects to infer the theoretical dark matter halo population described by Equation~\ref{eq:HMF} from the observed distribution of galaxy clusters. In its Bayesian framework, the observed cluster number counts are fit with an unbinned Poisson likelihood. In this section, we summarize the general framework and the modifications made to adapt the EDE models. 

\subsection{Statistical inference}
\label{sec:statistical_inference}

The distribution of observed cluster counts is modeled as a Poisson process with intensity $I(\chi|\Theta)$, where $\Theta$ denotes the set of cosmological parameters. The expected number of clusters with observables $\chi$ within the survey footprint $\Omega$ is obtained by integrating the intensity:

\begin{equation}
    N_{[\chi\in\Omega]}(\Theta)=\int_\Omega I(\chi|\Theta)\;\dd \chi.
\end{equation}

For our analysis, the vector of observables is expressed as: 

\begin{equation}
    \chi = \{\hat{C_\mathrm{R}},\hat{z},\hat{\mathcal{H}},\hat{\lambda},\hat{g_t}\},
\end{equation}
where the hat denotes the observed rather than the intrinsic quantity. In particular, $\hat{C_\mathrm{R}}$ is the observed X-ray count rate, $\hat{z}$ is the photometric redshift, $\hat{\mathcal{H}}$ is the sky position, $\hat{\lambda}$ is the optical richness and $\hat{g_t}$ is the WL tangential shear profile for the overlapping clusters. Given these observables, the intensity can be written as the differential number density in observable space:

\begin{equation}
    I(\chi|\Theta)=\frac{\dd \hat{N}}{\dd \hat{C}_\mathrm{R}\dd \hat{\lambda}\dd \hat{g_\mathrm{t}}\dd \hat{z}\dd \hat{\mathcal{H}}}.
\end{equation}

The connection between halos and observables is modeled as
\begin{equation}
    \frac{\dd\hat{N}}{\dd\hat{C}_\mathrm{R}\dd\hat{\lambda}\dd\hat{g_\mathrm{t}}\dd\hat{z}\dd\hat{\mathcal{H}}} = \int \mathcal{P}(\mathcal{I}|\chi)\,\mathcal{P}(\chi|M,z)\,\frac{\dd N}{\dd M}\,\frac{\dd V}{\dd z}\,\dd M\,\dd z,
\end{equation}
where $\mathcal{P}(\mathcal{I}|\chi)$ is the selection function described in Sec.~\ref{sec:selection_function}, and $\mathcal{P}(\chi|M,z)$ is the distribution of observables for a cluster of given mass and redshift (see Sec.~\ref{sec:scaling_relations}).

\subsection{Selection function}
\label{sec:selection_function}

The selection effects in the eRASS1 sample are modeled using high-fidelity end-to-end simulations with eROSITA's digital twin \citep{Seppi2021}. These simulations reproduce the survey’s cluster detection process, enabling an accurate evaluation of the selection function \citep{Clerc2024}. DM only UNIT1i simulations \citep{Chuang2019} are used to generate a full-sky light cone representing the large-scale structure, which is then populated with models of cluster and AGN X-ray emission \citep{Comparat2019, Comparat2020, Liu2022, Seppi2021}. To mimic eROSITA observations, synthetic X-ray events are generated and processed in the same way as real data, producing catalogs of detected sources. The probability of detecting a cluster is then estimated as a function of its intrinsic properties, accounting for survey properties, e.g., sky location, Galactic column density, and X-ray foreground and background. This probability is fully incorporated into the inference framework when modeling the survey's selection function as described in Section~\ref{sec:statistical_inference}.

For our particular analysis, we assume that $\mathcal{P}(\mathcal{I}|C_R,z,\hat{\mathcal{H}})$,  estimated via \lcdm simulations, is valid in the explored EDE cosmologies. Since EDE dissipates quickly after recombination and does not modify the late-time expansion, the angular diameter distance at low redshifts is well described by $\Lambda$CDM. As such, the cluster surface brightness for a given $C_R(<R_{500c})$ should remain unchanged in EDE cosmologies with $f_\mathrm{EDE}>0$ in comparison to the \lcdm simulations.

\subsection{Scaling relations}
\label{sec:scaling_relations}

The scaling relations between the observables used as mass proxies, the X-ray count rate $C_\mathrm{R}$ and the optical richness $\lambda$, and the underlying dark matter field are fit with varying cosmological parameters. The theoretical X-ray observable, $C_\mathrm{R}$, is assumed to follow \citetalias{Ghirardini2024}:

\begin{equation}
    \left<\ln\frac{\overline{C_\mathrm{R}}}{C_{R,p}}|M,z\right>=\ln A_\mathrm{X}+b_\mathrm{X}(M,z)\cdot \ln\frac{M}{M_p}+e_\mathrm{X}(z),
    \label{eq:CR_M}
\end{equation}
where $C_\mathrm{R,p}=0.1$~cts/s,  $M_p=2\cdot10^{14}M_\odot$, $z_p=0.35$ are pivot count-rate, mass, and redshift based on the sample median values, respectively. The redshift-dependent slope of the scaling relation is given by: 
\begin{equation}
    b_\mathrm{X}(z)=B_\mathrm{X}+F_\mathrm{X}\cdot \ln\frac{1+z}{1+z_\mathrm{p}},
\end{equation}
and the redshift evolution term, $e_\mathrm{X}(z)$, is:
\begin{equation}
    e_\mathrm{X}(z)=D_\mathrm{X}\cdot \ln \frac{d_\mathrm{L}(z)}{d_\mathrm{L}(z_\mathrm{p})}+E_\mathrm{X}\cdot \ln \frac{E(z)}{E(z_\mathrm{p})}+G_\mathrm{X}\cdot \ln \frac{1+z}{1+z_\mathrm{p}},
\end{equation}
The parameters $\{A_\mathrm{X}, B_\mathrm{X},  F_\mathrm{X}, G_\mathrm{X}\}$ are fitted simultaneously with the cosmological parameters, ensuring self-consistency and avoiding astrophysical biases, while $D_\mathrm{X}=-2$ and $E_\mathrm{X}=2$ are fixed according to the self-similar model \citep{Kaiser1986}.

Similarly, the optical richness $\lambda$ is modeled as:
\begin{equation}
    \left<\ln\lambda|M,z\right>=\ln A_\lambda +b_\lambda(z)\cdot\ln\frac{M}{M_\mathrm{p}}+C_\lambda\cdot\ln\frac{1+z}{1+z_\mathrm{p}},
\end{equation}
where the redshift dependence of the mass slope, $b_\lambda(z)$, follows:
\begin{equation}
    b_\lambda(z)=B_\lambda+D_\lambda\cdot\ln\frac{1+z}{1+z_\mathrm{p}}.
\end{equation}

The normalization $A_\lambda$, the redshift-dependent mass slope $B_\lambda$, the redshift evolution $C_\lambda$, and the evolution of the mass slope $D_\lambda$ are left to vary in the fits. 

\subsection{Weak lensing mass calibration}
\label{sec:weak_lensing_mass_callibration}

The use of the eRASS1 cluster catalog for cosmological inference requires the calibration of the mass-observable scaling relations presented in Section~\ref{sec:scaling_relations}. To do so, we use the overlapping DES \citep{Grandis2024a}, KiDS \citep{Kleinebreil2024}, and HSC \citep{Chiu2025} surveys. In the common region between eRASS1 and these surveys, each shear profile is directly injected in the mass calibration likelihood, as in \citetalias{Ghirardini2024}:
\begin{equation}
    \label{eq:mass_callibration}
    \mathcal{L}(\Theta) = \prod_{i}\mathcal{P}(\hat g_{\mathrm{t}}(R_i)|\chi_i,\Theta),
\end{equation}
where $g_\mathrm{t}(R_i)$ is the reduced tangential shear profile at the radius $R_i$. We compute this term by modeling the expected shear profile $g_\mathrm{t}(R,M_\mathrm{wl},z)$, where $M_\mathrm{wl}$ are the weak lensing estimated masses, which are known to be biased low when compared to the full dark matter mass $M$ as described in \cite{Grandis2024a}, and the difference should be estimated using numerical simulations. This bias is modeled as
\begin{equation}
\label{eq:wlbias}
\langle  \ln \frac{M_\mathrm{wl}}{M_\mathrm{p}}\bigg| M, z \rangle = b(z) + b_{M} \ln \left( \frac{M}{M_\mathrm{p}} \right),
\end{equation}
where the pivot mass is fixed to $M_\mathrm{p}=10^{14}M_\odot$, and $b(z)$ is the redshift evolution of the scatter, with the same parameterization as the one described in \cite{Ghirardini2024}.
The scatter of the weak lensing mass follows the same relation and reads
\begin{equation}
\label{eq:scatter_wl}
\ln \sigma_{M_\mathrm{wl}}^2 = s(z) + s_{M} \ln \left( \frac{M}{M_\mathrm{p}} \right) .
\end{equation}
\subsection{Mixture model}
\label{sec:mixtureModel}

The objects in the eRASS1 sample are divided into 3 classes: Clusters (C), which are our objects of interest for cosmological inference, and contaminants, which are AGNs or random fluctuations misclassified as clusters (NC). Through the Poisson mixture model, cluster abundances and contamination fractions are accounted for simultaneously. The total intensity of the Poisson process will be given by the sum of the three components:

\begin{equation}
    I_\mathrm{tot}(\chi|\Theta)=I_C(\chi|\Theta)+I_\mathrm{AGN}(\chi)+I_\mathrm{NC}(\chi)
\end{equation}
and the total number of objects will be given by:
\begin{equation}
    N_\mathrm{tot}(\Theta)=N_\mathrm{C}(\Theta)+f_\mathrm{AGN}N_\mathrm{tot}(\Theta)+f_\mathrm{NC}N_\mathrm{tot}(\Theta),
\end{equation}
where $f_\mathrm{AGN}$ and $f_\mathrm{NC}$ are the fractions of contaminants and are fitted simultaneously. With these we can express $N_\mathrm{tot}$, $N_\mathrm{NC}$, and $N_\mathrm{AGN}$ as a function of the cosmology-dependent $N_C(\Theta)$. With this, the total number of objects can be written as
\begin{equation}
    N_\mathrm{tot}(\Theta)=\frac{1}{1-f_\mathrm{AGN}-f_\mathrm{NC}}\int_\chi I_\mathrm{C}(\chi|\Theta)\;\dd \chi.
\end{equation}

The general shape of the Poisson log-likelihood will look like:
\begin{equation}
    \ln{\mathcal{L}(\Theta)=\sum_i\ln{I(\chi_i|\Theta)}}-\int_\chi I(\chi|\Theta)\;\dd \chi,
\end{equation}
where the index $i$ runs over the observed objects. Given a probability distribution function for the contaminants as a function of the observables, $\mathcal{P}_i(\chi)$, the number density of contaminants will follow $\lambda_\mathrm{AGN}=N_\mathrm{AGN}(\Theta)\mathcal{P}_\mathrm{AGN}(\chi)$ and $\lambda_\mathrm{NC}=N_\mathrm{NC}(\Theta)\mathcal{P}_\mathrm{NC}(\chi)$. With this formalism, the likelihood becomes:
\begin{equation}
\begin{aligned}
    \ln \mathcal{L}(\Theta) 
    &= \sum_i \ln \bigg[ 
        I_\mathrm{C}(\chi_i|\Theta) \\ 
    &\quad    + N_\mathrm{AGN}(\Theta)\,\mathcal{P}_\mathrm{AGN}(\chi_i)
    + N_\mathrm{NC}(\Theta)\,\mathcal{P}_\mathrm{NC}(\chi_i) 
    \bigg] \\
    &\quad - \frac{1}{1 - f_\mathrm{AGN} - f_\mathrm{NC}}
    \int_\chi I_\mathrm{C}(\chi|\Theta)\;\dd \chi .
\end{aligned}
\end{equation}
\subsection{Impact of EDE on the halo mass function}
\label{sec:hmf}
We now describe our theoretical model for the dark matter halo number density, including the adaptations made to model it in an EDE cosmology, and discuss its predictions for CMB-fitted cosmologies.

We define the root mean square of fluctuations filtered on a scale $R$, which encloses a corresponding mass $M=4\pi\rho_\mathrm{m}R^3/3$,  where $\rho_\mathrm{m}$ is the present-day mean matter density. It is given by:
\begin{equation}
    \sigma^2(M,z)\equiv\frac{1}{2\pi^2}\int k^2W^2(kR)P_\mathrm{lin}(k,z)\;\dd k,
\label{eq:sigma}
\end{equation}
where $W(y)= 3[\sin(y)/y^3-\cos(y)/y^2]$ is the Fourier transform of the top-hat window function in the real-space of radius R. 

Following the Press-Schechter formalism \citep{PressSchechter1974}, the HMF will be given by 
\begin{equation}
    \frac{\dd n}{\dd\ln M}=\frac{\rho_\mathrm{m}}{M}\left|\frac{\dd\ln\sigma}{\dd\ln M}\right|f(\sigma),
\label{eq:HMF}
\end{equation}
where the multiplicity function $f(\sigma)$ relates the effects of non-linear collapse to $\sigma(M,z)$. The fitting function used in this work is \cite{Tinker2008}; tests on the effects of using the \cite{Despali2016} or including additional nuisance parameters to the HMF are provided in App.~\ref {sec:HMF_Systematics}. 

We use the \texttt{EarlyQuintessence} class developed by \citet{Smith2020} for the Boltzmann solver \texttt{CAMB} \citep{Lewis2000} to generate the EDE power spectrum, and use the \texttt{Colossus} package \citep{Diemer2018} to compute the HMF using the multiplicity function from \citet{Tinker2008}. For our baseline analysis, as is common in the literature, we fix $n=3$. This choice
has been shown to fit the current data and remains close to the best-fit value when left as a free parameter \citep{Poulin2019, Smith2020}.
This is further motivated, as argued in \cite{Hill2020}, by the theoretical argument that the phenomenological model corresponds to a careful fine-tuning of the transition hierarchy where, for a specific integer $n$, only the first integers up to $n$ require fine-tuning. This especially favors $n=2$ and $n=3$, as they require minimal fine-tuning while ensuring that the field decays more rapidly than matter once it starts to oscillate. As a test of the sensitivity of our results to this choice, we perform an analysis with $n$ left free to vary; the results are shown in App.~\ref{sec:exponent_tests}.

\subsubsection{Validation of fitting function}
\label{sec:validation}

Due to the high computational cost, a dedicated calibration of $f(\sigma)$ has not yet been performed for EDE cosmologies. However, since EDE couples only weakly to the metric and rapidly dissipates after recombination, it does not introduce new interactions with standard-model species during nonlinear collapse. Therefore, especially in the galaxy cluster mass range where non-linear collapse effects are less dominant, its effect on the HMF should arise mainly through the modified growth history encoded by shifts in the equivalent \lcdm parameters, with the linear effects captured by $\sigma(M,z)$. Following this argument, the implications for non-linear effects in the galaxy cluster mass range are assumed to remain captured through the \lcdm calibration of $f(\sigma)$ by \citet{Tinker2008}, within an estimated error budget estimated of $5-10\%$ \citep{Artis2021}. We test this assumption by comparing our HMF model with the \cite{Klypin2021} EDE cosmology N-body simulations, which use the best-fit parameters reported in \cite{Smith2020}. Using the simulation results, we construct the halo abundance $\dd N/\dd\ln M$ and compare it to our predictions.  

\begin{figure}[htb]
\centering
    \includegraphics[width=\linewidth]{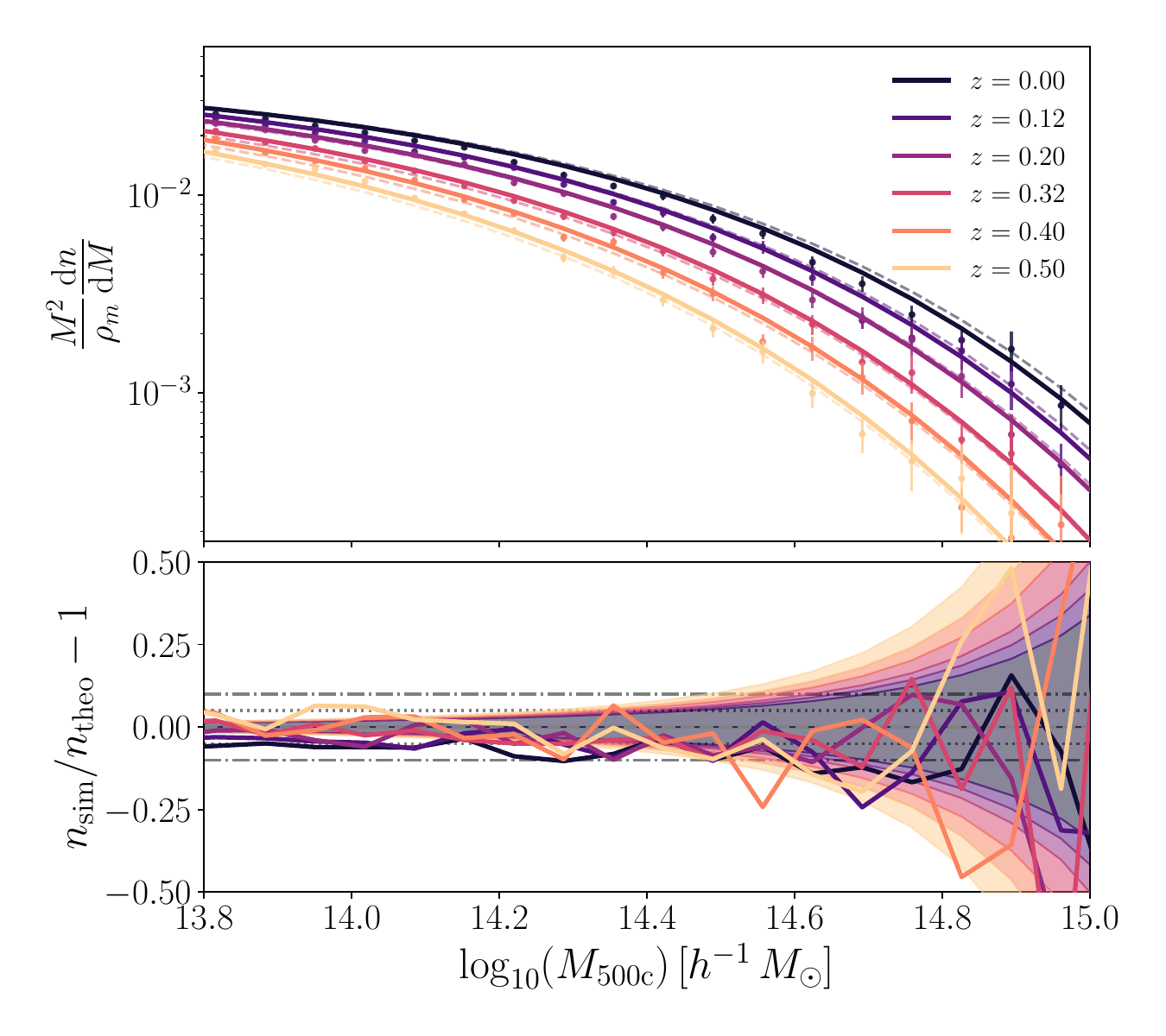}
    \caption{Halo mass function ($M_{500\mathrm{c}}$) at different redshifts. Solid lines represent the HMF for an EDE cosmology derived through Equation~\ref{eq:HMF} using the \citet{Tinker2008} multiplicity function, for the best-fit parameters reported in \citet{Smith2020}. Points with error bars show the binned EDE halo abundances computed from the  \citet{Klypin2021} N-Body Simulations for the same cosmology, with their associated Poisson uncertainty. Dashed lines represent the \lcdm HMF model using the \citet{Tinker2008} multiplicity function, evaluated at the \planck best-fit parameters. The bottom panel shows the relative difference between the predicted cluster abundances from our model and those observed in the simulations. The gray dashed horizontal lines represent the 5 and 10 \% error margins for the \citet{Tinker2008} multiplicity function \citep[as estimated in][]{Artis2021}. Filled areas represent the associated theoretical Poisson error $\sim1/\sqrt{n_\mathrm{theo}}$.}
    \label{fig:Fig2}
\end{figure}

As can be seen in the lower plot of Fig.~\ref{fig:Fig2}, our theoretical model reproduces the cluster population in the $10^{13.8}-10^{15}\;h^{-1}M_\odot $ range of the \citet{Klypin2021} simulation, within the redshift range $0.1<z<0.45$ of the eRASS1 local cluster catalog. At the lower mass range, our model reproduces the population within the $f(\sigma)$ calibration error budget, and at the higher mass range within the Poisson error envelope $\sim1/\sqrt{n_\mathrm{theo}}$.

To ensure our results are robust against HMF calibration choices, we perform additional tests and show the results in App.~\ref{sec:HMF_Systematics}. We test the impact of systematic uncertainties on the halo mass function by following the formalism of \citet{Costanzi2019}, which introduces nuisance parameters for the HMF normalization and slope. Additionally, we examine the constraints obtained when using the \citet{Despali2016} multiplicity-fitting function rather than \citet{Tinker2008}. We find good agreement in the constraints obtained regardless of changes to the fitting function or the addition of systematics of the order of 5\%, highlighting that changes in the mass function do not significantly impact our constraints with the current data. 

Given that our cosmological constraints are largely insensitive to the tested variations, and that our theoretical model reproduces N-body simulation abundances within the budget required for eRASS1 precision cosmology, we consider it justified for our analysis to assume that the \citet{Tinker2008} $f(\sigma)$ reproduces EDE cluster abundances in the studied mass and redshift range. We note, however, that this validation was performed only for one EDE cosmological realization due to the limited availability of such simulations. 

\subsubsection{Impact of Hubble tension-resolving EDE on the halo mass function}
\label{sec:power_spectrum_impact}

Motivated by the good agreement between our model and the available simulations, we use it to study the EDE halo mass function across parameter spaces that can resolve the Hubble tension.
For this purpose, we adopt a fiducial cosmology based on the best-fit parameters reported by \citet{Smith2020}, which were constrained using primary CMB data, BAO, RSD, and \citetalias{Riess2019}-calibrated SN~Ia.

As briefly discussed in Sec.~\ref{sec:intro}, the primary direct effect of EDE on the matter power spectrum is the suppression of the growth of perturbations inside the horizon, during the period where the EDE-induced boost is relevant to the expansion rate. As such, the critical redshift $z_\mathrm{c}$ regulates the horizon scale $k_\mathrm{c}$, such that subhorizon modes with $k\gtrsim k_c$ have their growth suppressed. The overall strength of the suppression is determined by $f_\mathrm{EDE}$, where an increase further suppresses the growth, and predominantly impacts the high-$k$ modes. 
To maintain the fit to early-Universe observations, this EDE-induced suppression is offset through shifts in the standard \lcdm parameters, primarily encapsulated by increases in $\Omega_\mathrm{m}h^2$, $n_\mathrm{s}$, and $A_\mathrm{s}$. These are necessary to compensate for the early Integrated Sachs-Wolfe effect in the CMB (eISW), where the EDE expansion boost causes the gravitational potential to decay more rapidly \citep{Poulin2019, Hill2020, Vagnozzi2021}, and to keep the BAO scale fixed at late times \citep{Jedamzik2021}.
Additionally, because the CMB tightly constrains the physical densities $\omega_\mathrm{m,b}=\Omega_\mathrm{m,b} h^2$, EDE cosmologies with a higher $H_\mathrm{0}$ typically require smaller values of $\Omega_\mathrm{m,b}$. The model increases the cosmological constant to improve the fit to low-redshift SN~Ia data \citep{GomezValent2022}, leading to an earlier onset of Dark Energy (DE) domination and its corresponding late-time suppression of structure growth.

\begin{figure}[htb]
\centering
    \includegraphics[width=\linewidth]{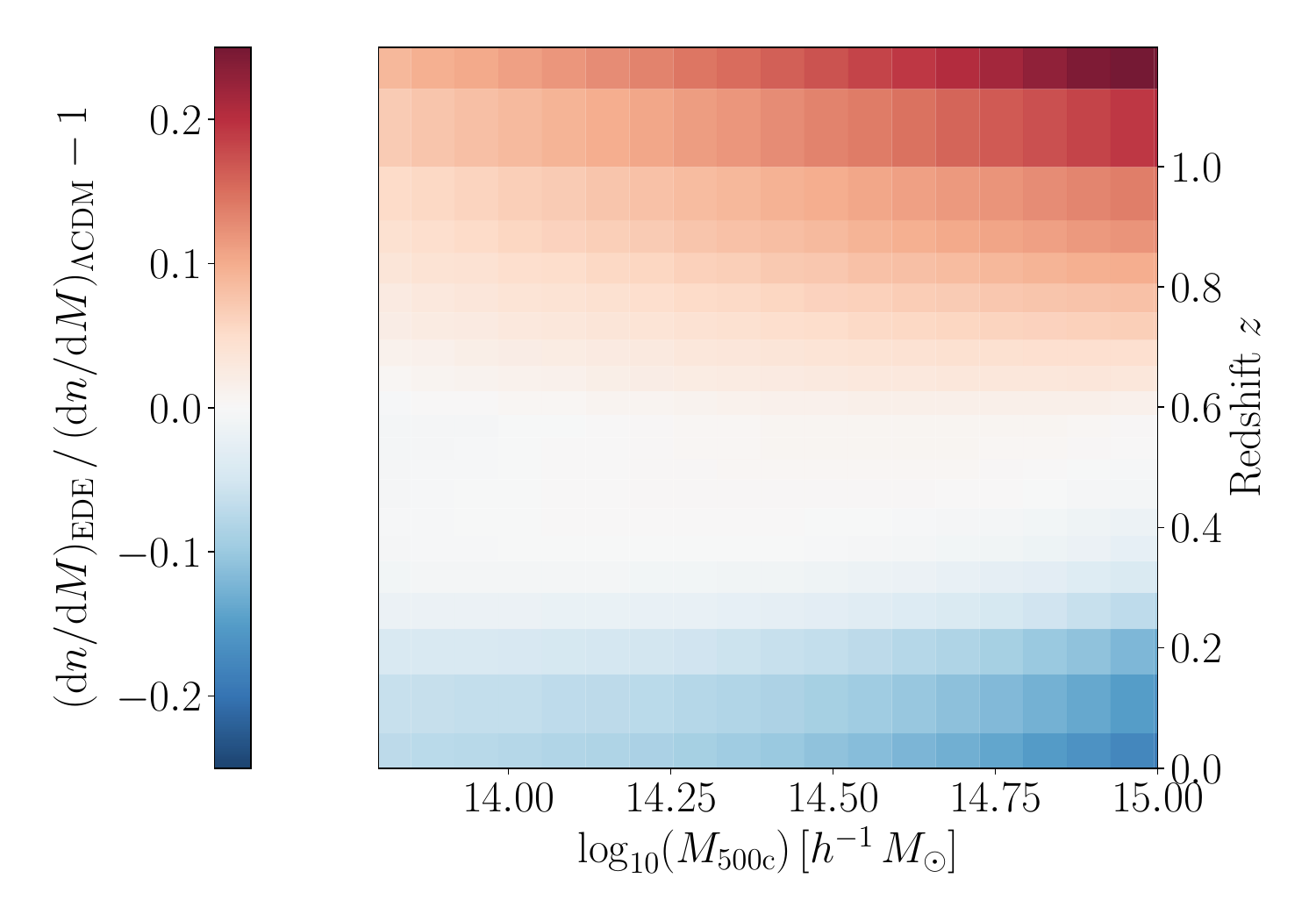}
    \caption{Relative differences between the halo mass function for an EDE cosmology based on the \citet{Smith2020} best-fit parameters and for a \planck \lcdm cosmology \citep{PlanckCollaboration2020d}, both derived through Equation~\ref{eq:HMF} using the \citet{Tinker2008} multiplicity function.}
    \label{fig:Fig3}
\end{figure}

We illustrate the consequences of these shifts on the HMF in Fig.~\ref{fig:Fig3}, comparing our fiducial EDE evaluated at the best-fit parameters reported in \citet{Smith2020} against a \planck \lcdm cosmology.
The modifications of the power spectrum are reflected in the HMF, characterized by a higher $\sigma_\mathrm{8}$ and a lower $\Omega_\mathrm{m}$. At higher redshifts, the HMF predicted by EDE shows more clusters, with the relative difference increasing with cluster mass. Conversely, at lower redshifts, the HMF model for the fiducial EDE cosmology predicts fewer clusters across all mass ranges than in \lcdm. Despite the higher $\sigma_\mathrm{8}$, the enhanced clustering at these redshifts is balanced out by the lower $\Omega_\mathrm{m}$, which implies an earlier dark energy domination compared to \lcdm. For $z\lesssim0.3$, this leads to an overall minor suppression of the cluster population with decreasing redshift, with the effect being further pronounced at the massive cluster tail. 

\section{Results}
\label{sec:results}
This section presents the results of a Bayesian cosmological analysis of an EDE cosmology, based on galaxy cluster counts.
We first present the constraints derived from eRASS1 cluster number counts and compare them with the literature. This is followed by the results obtained when combining our analysis with the posteriors reported in \citetalias{Calabrese2025}, containing CMB anisotropy measurements from \citetalias{Louis2025} and  \citetalias{PlanckCollaboration2020}, CMB lensing measurements from ACT~DR6 \citep{Qu2024,Madhavacheril2024} and \planck \citep{PlanckCollaboration2020b}, and BAO measurements from DESI \citep{Adame2025, Qu2024b}. 
We run all our chains using the \texttt{emcee} sampler \citep{emcee2013} until the Gelman-Rubin \citep{Gelman1992} convergence criterion $|\hat{R}-1|<0.01$ is reached.

\begin{table*}[htb]
\caption{Priors used on the cosmological parameters.}
 \centering
 \begin{tabular}{lll}
 \hline
 \hline
 Parameter  & Description & Prior \\
 \hline
 $\Omega_{\mathrm{m}}$ & Mean matter density at present time & $\mathcal{U}(0.16, 0.5)$ \\
 $\log_{10} A_\mathrm{s}$  & Amplitude of the primordial power spectrum & $\mathcal{U}(-9, -8)$ \\
 $H_\mathrm{0}$ $[{\rm km}\,{\rm s}^{-1}\,{\rm Mpc}^{-1}]$  & Hubble expansion rate at present time & $\mathcal{U}(60, 80)$ \\
 $\Omega_{\mathrm{b}}$ & Mean baryon density at present time & $\mathcal{U}(0.04, 0.06)$ \\
 $n_\mathrm{s}$  & Spectral index of the primordial power spectrum & $\mathcal{U}(0.92, 1.05)$  \\
 $f_\mathrm{EDE}$  & Maximal fractional EDE contribution & $\mathcal{U}(0.0001, 0.5)$\\
 $\log_{10}z_\mathrm{c}$  & Redshift & $\mathcal{U}(2.5, 4.5)$\\
 $\theta_\mathrm{i}$  & Initial field value & $\mathcal{U}(0, 3.1)$\\
 \hline
 \hline
 \end{tabular}
  \tablefoot{For the full set of priors used for the other parameters, including scaling relation and nuisance parameters, see \citetalias{Ghirardini2024}. $\mathcal{U}(a,b)$ denotes a uniform distribution between $a$ and $b$. 
  }
 \label{tab:parameters_priors}
\end{table*}

\subsection{Constraints from eRASS1 cluster abundances}
\label{subsec:erosita_results}

To obtain constraints on EDE using cluster counts, we assume a flat $\Lambda\mathrm{CDM}$ cosmology with an EDE extension. We fit the cluster number density to the eRASS1 cosmology sample, leaving the parameters $\Omega_\mathrm{m}$, $\Omega_\mathrm{b}$, $n_\mathrm{s}$, and $\log A_\mathrm{s}$, and $H_\mathrm{0}$ free with the priors used in the G24 cosmology analysis. Additionally, the EDE model includes parameters $f_\mathrm{EDE}$, $\log z_\mathrm{c}$ and $\theta_\mathrm{i}$, as indicated in Tab.~\ref{tab:parameters_priors}. To ensure a fair comparison with CMB analyses, we similarly assume a fixed neutrino mass sum of $\sum{m_\nu}=0.06\; \mathrm{eV}$. The inclusion of massive neutrinos in the analysis may affect the halo mass function, as described in detail in \citetalias{Ghirardini2024}. We account for this effect by following the prescription described in \citet{Costanzi2013}. We note that while in the \lcdm eRASS1-only analysis of \citetalias{Ghirardini2024} $\sum{m_\nu}=0\; \mathrm{eV}$ is left fixed, as discussed in their Appendix B.4 and reproduced for the EDE case, the constraints obtained by fixing $\sum{m_\nu}=0.06$ are only marginally modified.

\begin{figure*}[htb]
\sidecaption
    \includegraphics[width=12cm]{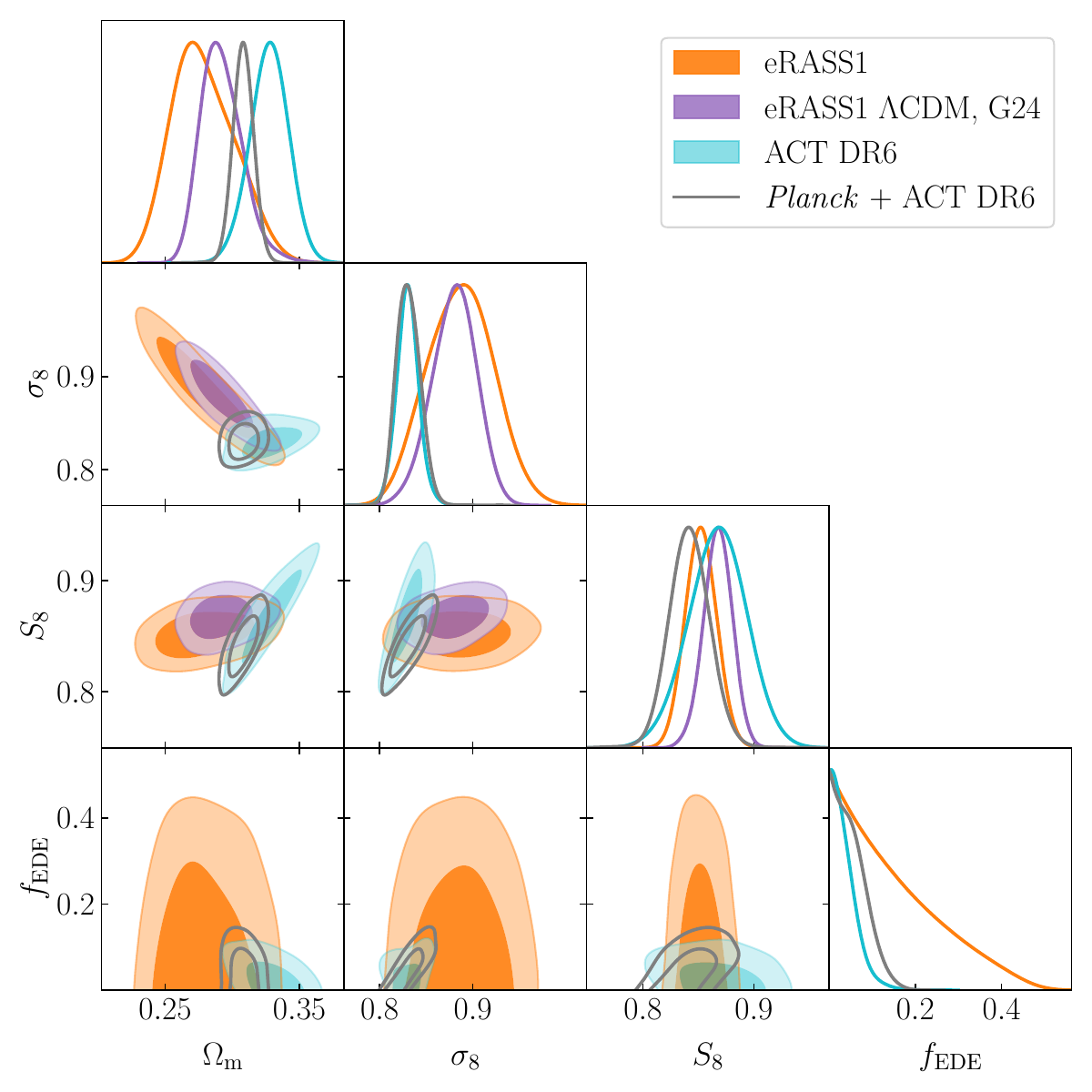}
    \caption{Marginalized posteriors for the parameters $\Omega_\mathrm{m}$, $\sigma_\mathrm{8}$, $S_8$ and $f_\mathrm{EDE}$ constrained in a flat \lcdm model with an EDE extension for the eRASS1 $z<0.45$ analysis, in orange filled contours. For comparison, we show the posteriors reported in \citetalias{Calabrese2025}, including \citetalias{Louis2025} with a prior on $\tau_\mathrm{reio}$ (in cyan-filled contours), and the ACT data combined with \citetalias{PlanckCollaboration2020} (in gray solid lines). We also show the constraints from the eRASS1 \lcdm analysis published in \citetalias{Ghirardini2024}, in filled purple contours.}
    \label{fig:Fig4}
\end{figure*}

For the eRASS1 cluster count analysis of EDE, we find $\Omega_\mathrm{m}=\OmegameRASS$, $\sigma_\mathrm{8}=\sigmaEighteRASS$, and $S_8=\SheRASS$ \noindent at 68\% confidence level. The baseline \lcdm analysis presented in \citetalias{Ghirardini2024} with the best-fit values of $\Omega_\mathrm{m}=0.29_{-0.02}^{+0.01}$, $\sigma_\mathrm{8}= 0.88\pm0.02$, and $S_8=0.86\pm0.01$,  are consistent with our findings at the $1\sigma$ confidence level (see Table~\ref{tab:all_constraints}). The posterior parameters and comparisons with the \citetalias{Ghirardini2024} analysis are shown in Figure~\ref{fig:Fig4}. The consistency with the standard cosmology analysis further confirms the robustness of our flat \lcdm analysis. 

We obtain an upper limit on the EDE fraction of $f_\mathrm{EDE}<\fEDEeRASS$ at the 95\% confidence level. This constraint is not stringent enough to exclude an EDE fraction of $f_{\mathrm{EDE}}\sim0.1$, which has been shown to alleviate the Hubble tension \citep{Poulin2018, Smith2021}. The inclusion of EDE in our analysis marginally decreases our constraining power on the \lcdm parameters in comparison to \citetalias{Ghirardini2024}, which is exacerbated by our use of a reduced redshift sample with $z<0.45$. Our at-large agreement with the \lcdm analysis shows that cluster counts are not sensitive to the early-time effects of EDE. As such our analysis offers only marginal constraining power on $\log_{10}(z_\mathrm{c})=\logzceRASS$, and none on the $\theta_i$ parameter.

Our constraints on $S_8$, $\sigma_8$, and $\Om$ are in good agreement with EDE constraints obtained from CMB measurements, as shown in Fig.~\ref{fig:Fig4}. Our results agree 
within $\sim1.7\sigma$ with constraints reported from the \citetalias{PlanckCollaboration2020}-only analyses in the literature \citep{Hill2020, McDonough2024, Qu2024b}, the ACT DR6 data within $\sim1.2\sigma$ \citepalias[see \autoref{subsec:CMB_combination} for further discussion]{Calabrese2025} and the SPT-3G D1 data within $\sim1.5\sigma$ \citep{Khalife2025}. This consistency remains within $\sim1.7\sigma$ of measurements that also include CMB lensing \citep{Qu2024b}. These CMB analyses report upper limit constraints at 95\% confidence of $f_\mathrm{EDE}<0.1$ to different degrees of precision, with the critical redshift placed around $\logzc\sim3.5$, with which our EDE constraints are fully compatible. 

The cluster halo mass function alone is insensitive to the $H_\mathrm{0}$ parameter due to its large degeneracy with the other cosmological parameters, and thus we cannot constrain it to the precision comparable with direct or early-time measurements, as discussed in \citetalias{Ghirardini2024}. Within the \lcdm framework and in extensions that modify the expansion history, a common strategy has been to adopt more informative Gaussian priors to break the degeneracy between the expansion history and the growth of structure. This is done with a prior either centered around CMB-inferred indirect measurements (as in \citetalias{Ghirardini2024}), a value informed by late-time measurements \citep[e.g. one obtained with calibrated SN~Ia data as in][]{deHaan2016}), or a prior wide enough to encompass all current high-precision constraints \citep[e.g. $H_\mathrm{0}\sim\mathcal{N}(70, 5)$ as in][]{Bocquet2024}. Cluster abundance analyses in \lcdm provide robust constraints on $\Om$, $\sigma_{8}$, and $S_8$ and that are largely insensitive to the choice of 
$H_\mathrm{0}$ prior \citep{deHaan2016, Ghirardini2024}. Motivated by this, we examine the effect of varying the $H_\mathrm{0}$ prior on our EDE constraints, as discussed in App.~\ref{sec:H0_priors}.
Crucially, as expected given the insensitivity of cluster counts to $H_\mathrm{0}$ within the range favored by geometric probes, we find that the constraints on $S_8$ as well as $f_{\mathrm{EDE}}$ are consistent independently of the $H_\mathrm{0}$ prior used. 

The $f_{\mathrm{EDE}}$ upper limits effectively constrain the maximal EDE-induced suppression on structure growth compatible with eRASS1 cluster abundances. Since our constraints on $S_8$ and the EDE parameters constitute an independent analysis, they can, if compatible, be jointly analyzed with cosmological probes sensitive to the early-time effects of EDE on the expansion rate. To assess the compatibility of our analysis with early-Universe probes and the validity of a joint analysis, we quote and present the results we obtain with a flat prior on $H_\mathrm{0}\sim\mathcal{U}(60,80)$, thereby avoiding the bias introduced by the assumption of Gaussian priors from geometric probes.

\begin{table*}[htb]
\caption{Marginalized constraints on the baseline cosmological and EDE parameters.}
\centering
\begin{tabular}{lcccc}
\hline
\hline
Parameter & eRASS1 & + ACT & + P + ACT & + P + ACT + L + B\\
\hline
$\Omega_\mathrm{m}$ & $\OmegameRASS$ & $\OmegameRASSACT$ & $\OmegameRASSPACT$ & $\OmegameRASSPACTLB$ \\
$\sigma_\mathrm{8}$ & $\sigmaEighteRASS$ & $\sigmaEighteRASSACT$ & $\sigmaEighteRASSPACT$ & $\sigmaEighteRASSPACTLB$ \\
$S_8$ & $\SheRASS$ & $\SheRASSACT$ & $\SheRASSPACT$ & $\SheRASSPACTLB$ \\
$H_\mathrm{0}\;[{\rm km}\,{\rm s}^{-1}\,{\rm Mpc}^{-1}]$ & - & $\HZeroeRASSACT$ & $\HZeroeRASSPACT$ & $\HZeroeRASSPACTLB$ \\
$f_{\mathrm{EDE}}$ & $<\fEDEeRASS$ & $<\fEDEeRASSACT$ & $\fEDEeRASSPACT$ & $\fEDEeRASSPACTLB$ \\
$\logzc$ & $\logzceRASS$ & $\logzceRASSACT$ & $\logzceRASSPACT$ & $\logzceRASSPACTLB$ \\
\hline
\hline
\end{tabular}
\tablefoot{Constraints derived from the eRASS1 cluster abundance analysis ($0.1 < z < 0.45$) and its subsequent combination with posteriors from external datasets. External probe combinations include primary CMB data from ACT and \planck (P), CMB lensing (L), and DESI BAO (B). Uncertainties represent 68\% confidence intervals, while upper limits on $f_{\mathrm{EDE}}$ are quoted at the 95\% confidence level.
}
\label{tab:all_constraints}
\end{table*}
\subsection{Combination with CMB probes}
\label{subsec:CMB_combination}

In this section, we assess the consistency between our eRASS1 results and the EDE constraints derived from the early-time probe combination reported by \citet{Calabrese2025}. 
To quantify the agreement between our constraints and the \citet{Calabrese2025} probe combination results, we construct the posterior distribution of the parameter differences. We first compute the probability to exceed (PTE), which is defined as the integral over the region where the difference distribution is lower than at the origin, and thus provides a test of the null hypothesis that the difference between the measurements is zero. The PTE is then converted to a significance in confidence level, $\sigma$, by calculating the Gaussian equivalent. If a good agreement within $<2\sigma$ is reached, we then proceed with a combined analysis of our posteriors from the eRASS1 analysis with those from \citetalias{PlanckCollaboration2020} and \citet{Louis2025} CMB data (reported in \citetalias{Calabrese2025}) to reduce degeneracies and constrain the properties of the EDE model. 

\begin{figure*}[htb]
\sidecaption
    \includegraphics[width=12cm]{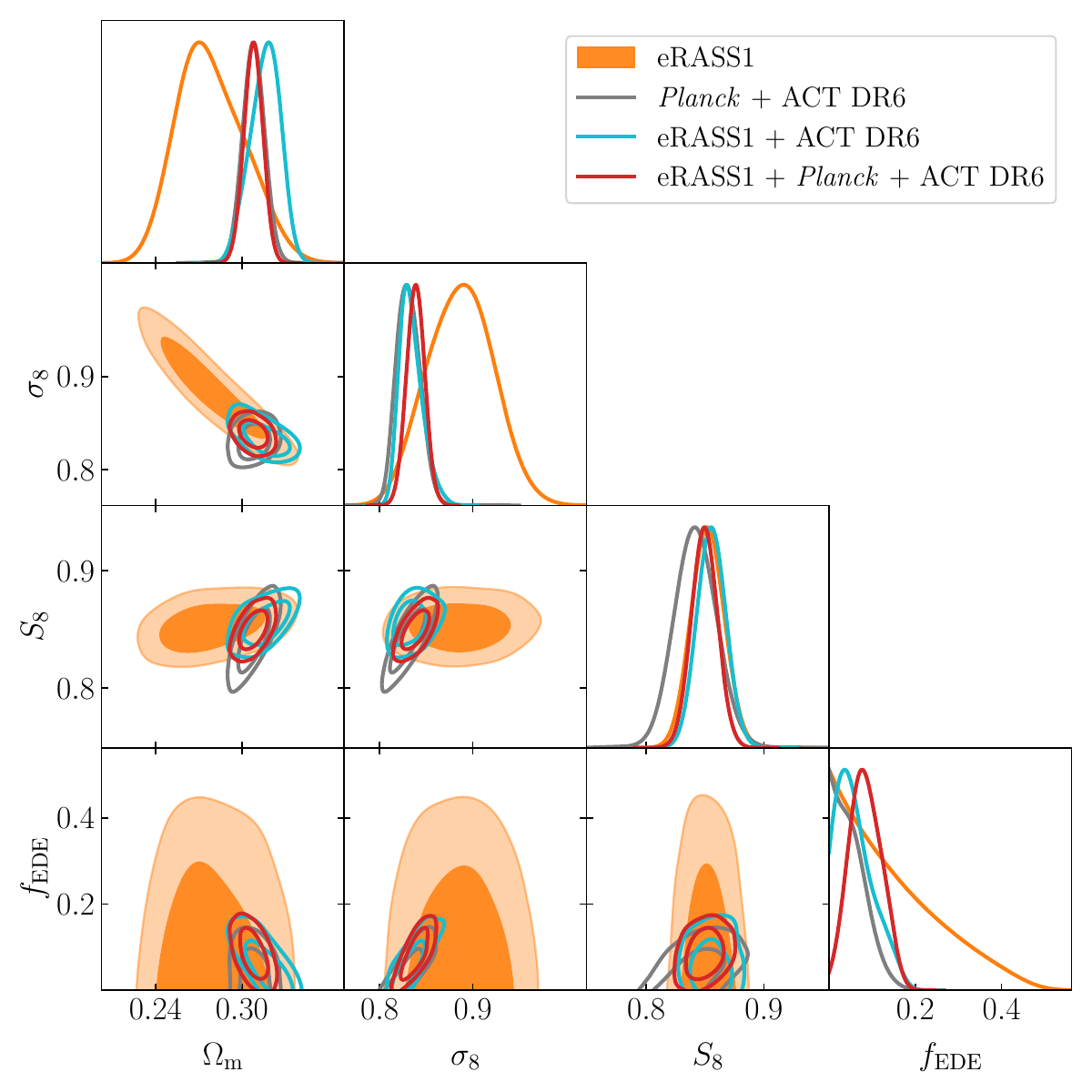}
    \caption{Marginalized posteriors for the parameters $\Omega_\mathrm{m}$, $\sigma_\mathrm{8}$, $S_8$ and $f_\mathrm{EDE}$ constrained in a flat \lcdm model with an EDE extension for the eRASS1 analysis combined with CMB data. In orange, we show the eRASS1-only analysis. The figure includes the EDE results published in \citetalias{Calabrese2025}, for \citetalias{PlanckCollaboration2020} and \citetalias{Louis2025}, in dark gray filled contours. We show in cyan lines the results for eRASS1 combined with ACT DR6 data, and in solid red lines the results for the joint eRASS1, \planck, and ACT analysis. Combined \planck, ACT, and eROSITA cluster number counts produce a $2.8\sigma$ detection of the EDE fraction of $f_\mathrm{EDE}=0.09\pm0.04$ and Hubble parameter of $70.3^{+1.3}_{-1.5}$~km~s$^{-1}$~Mpc$^{-1}$.} 
    \label{fig:Fig5}
\end{figure*}

We first compare our analysis with the EDE constraints obtained using ACT primary CMB measurements with a \citetalias{PlanckCollaboration2020} prior on $\tau_\mathrm{reio}\sim\mathcal{N}(5.66 \times 10^{-2}, \, 5.8 \times 10^{-3})$. This data yields strong constraints on the $\sigma_8$ parameter with a degeneracy direction with $\Om$ orthogonal to the one observed in our eRASS1 only analysis presented in Sec.~\ref {subsec:erosita_results}. 
We find that the PTE for $S_8$ is $0.58(0.5\sigma)$, and $0.23(1.2\sigma)$ for $\sigma_8$ and $\Om$, as well as EDE parameters (given in Table~\ref{tab:all_constraints}) consistent with the ACT results within 1$\sigma$ confidence level. 
We then proceed to combine our analysis with the posterior distributions from ACT. This combination enhances the sensitivity of our analysis to early-time EDE–induced modifications to the expansion history and yields tighter constraints by breaking the degeneracy between $\Omega_\mathrm{m}$ and $\sigma_\mathrm{8}$. The results of the joint analysis are shown in Tab.~\ref{tab:all_constraints} and Fig.~\ref{fig:Fig5}. Compared to the eRASS1-only results, we see a shift to a higher value of $\Om$ and to a lower value for $\sigma_8$, with the joint analysis yielding constraints corresponding to a factor of $\sim2$ improvement on our constraining power on $\Omega_\mathrm{m}$, and a factor of $\sim3$ on $\sigma_\mathrm{8}$ and $f_\mathrm{EDE}$.
While the joint constraint on $f_\mathrm{EDE}$ is still consistent with zero, the marginalized posterior is notably shifted towards higher values compared to the ACT constraints. Driven by the positive $f_\mathrm{EDE}$ and $H_\mathrm{0}$ degeneracy direction introduced by the ACT prior, this shift leads to a larger inferred value of $H_\mathrm{0}=\HZeroeRASSACT\,{\rm km}\,{\rm s}^{-1}\,{\rm Mpc}^{-1}$, reducing the tension with \citetalias{Riess2022} SN~Ia results to $0.09
(1.7\sigma)$ confidence level.

The agreement of eRASS1 with early-Universe probes is improved when compared to the combined \planck and ACT posteriors, where we find the PTE for $S_8$ to be $0.68\,(0.4\sigma)$, and $0.33\,(0.9\sigma)$ for $\sigma_8$ and $\Om$. Given this good agreement, we combine our eRASS1 analysis with the \planck and ACT probe combination \citepalias{Calabrese2025}, with the results shown in Tab.~\ref{tab:all_constraints} and Fig.~\ref{fig:Fig5}. The addition of \planck complements ACT with low-$\ell$ CMB measurements, improving constraints on $\Omega_\mathrm{m}$ through the eISW effect. This joint CMB analysis yields tighter constraints on the structure growth parameters, prefers a lower value of $S_8$ and introduces a positive correlation between $S_8$ and $f_\mathrm{EDE}$, in contrast to the almost orthogonal degeneracy direction seen for ACT alone. The combination with eRASS1 reconciles the datasets along this tilted degeneracy direction, with the higher $S_8$ from eRASS1 leading to a joint posterior that favors $f_\mathrm{EDE}>0$ at $0.005\,(2.8\sigma)$. The constraints obtained from eRASS1, \planck, and ACT are the first non-zero ($>2\sigma$) $f_\mathrm{EDE}$ constraints based solely on a Bayesian analysis of CMB and growth of structure, without including information from the cosmic distance calibration. 

Remarkably, through the degeneracy between $f_\mathrm{EDE}$ and $H_\mathrm{0}$ inherent in the CMB data, the combination of cluster abundances with primary CMB posteriors leads to inferred constraints on $H_\mathrm{0}=\HZeroeRASSPACT$~${\rm km}\,{\rm s}^{-1}\,{\rm Mpc}^{-1}$, largely consistent with the model-independent late-time expansion measurements. Our constraints are consistent within $0.11(1.5\sigma)$ of the \citetalias{Riess2022} direct measurements of $H_\mathrm{0}=73.04\pm1.04$~${\rm km}\,{\rm s}^{-1}\,{\rm Mpc}^{-1}$, which presents the strongest reported constraints for a single late-time geometric probes,
and are fully compatible within $\sim1\sigma$ with other recent high-precision measurements of $H_\mathrm{0}$. These include measurements based on the tip of the red giant branch (TRGB) calibration of Type Ia supernovae, yielding $H_\mathrm{0} = 70.39\pm1.22 \,(\mathrm{stat}) \pm 1.33 \,(\mathrm{sys}) \pm 0.70 \,(\sigma_\mathrm{SN})\,{\rm km}\,{\rm s}^{-1}\,{\rm Mpc}^{-1}$ \citep{Freedman2025}, those inferred from Mira variables, $H_\mathrm{0} = 73.06\pm2.67{\rm km}\,{\rm s}^{-1}\,{\rm Mpc}^{-1}$ \citep{Bhardwaj2025}, strong-lensing time-delay analyses, $ H_\mathrm{0} = 71.6_{-3.3}^{+3.9}{\rm km}\,{\rm s}^{-1}\,{\rm Mpc}^{-1}$ \citep{Birrer2025}, and single-step megamaser measurements, $H_\mathrm{0} = 73.9 \pm 3.0 {\rm km}\,{\rm s}^{-1}\,{\rm Mpc}^{-1}$ \citep{Pesce2020}.

Our constraints are consistent with the joint EDE analyses of CMB,  BAO, and galaxy clustering data in the literature, although we find a higher value of $\sigma_8$ and accordingly $S_8$, leading to higher, non-zero values of $f_\mathrm{EDE}$, in contrast to upper limit constraints. Our results agree within $2\sigma$ with the constraints reported by \cite{Ivanov2020} with CMB data combined with BOSS full-spectrum and BAO, $\Om=0.308\pm0.005$, $\sigma_8=0.813^{+0.007}_{-0.009}$, $S_8=0.824\pm0.011$ and $f_\mathrm{EDE}<0.072$, as well as those reported in \citet{Qu2024b} for DESI BAO combined with \citetalias{PlanckCollaboration2020} primary CMB and lensing data, $\Om=0.303^{+0.006}_{-0.004}$, $\sigma_8=0.822^{+0.007}_{-0.009}$, $S_8=0.826\pm0.010$, and $f_\mathrm{EDE}<0.091$. The joint analysis of the latter reported in \citetalias{Calabrese2025} using \citetalias{Louis2025} agrees within $\sim1.5\sigma$ with our results, and is discussed further in \autoref{subsec:CMB_LSS_combination}.

However, as discussed in Sec.~\ref{sec:intro},  the constraints placed on EDE by LSS remain under debate, potentially explaining the differences between our findings and previously reported constraints on $f_\mathrm{EDE}$. The debate centers particularly on the BOSS full-spectrum re-analyses that relax the upper limit constraints on $f_\mathrm{EDE}$ \citep{Simon2023,Holm2023}, and, more generally, on the susceptibility of EDE models to prior volume effects, which have been argued to bias Bayesian marginalized constraints towards $f_\mathrm{EDE}\sim0$ \citep{Niedermann2020, Murgia2021}. In this limit, the EDE model yields observables well described by an \lcdm cosmology. Consequently, $z_\mathrm{c}$ and $\theta_i$ lose phenomenological impact and are unconstrained. As such, even if regions with $f_\mathrm{EDE}>0$ fit the data better or equally well, the fine-tuned region of $\{z_\mathrm{c},\theta_i\}$ will occupy a comparatively smaller volume than the unconstrained \lcdm-like region around $f_\mathrm{EDE}\sim0$. This can dominate the marginalization process, resulting in upper limits that mask the model's ability to alleviate the tension \citep{Smith2021, GomezValent2022}. We discuss the potential impact of prior volume effects in our analysis in App.~\ref{sec:gof}.
The EDE results from combining eRASS1 and CMB posteriors agree with profile-likelihood constraints on $f_\mathrm{EDE}$, which, unlike Bayesian analyses, take a frequentist approach and circumvent prior volume effects by construction. Using this methodology, \citet{Herold2023} report comparable results, with $f_\mathrm{EDE}=0.072\pm0.039$ for \planck data and $f_\mathrm{EDE}=0.087\pm0.037$ for the further combination with the BOSS full-shape likelihood.

\subsection{Combination with CMB lensing and BAO}
\label{subsec:CMB_LSS_combination}
\begin{figure*}[htb]
\sidecaption
    \includegraphics[width=12cm]{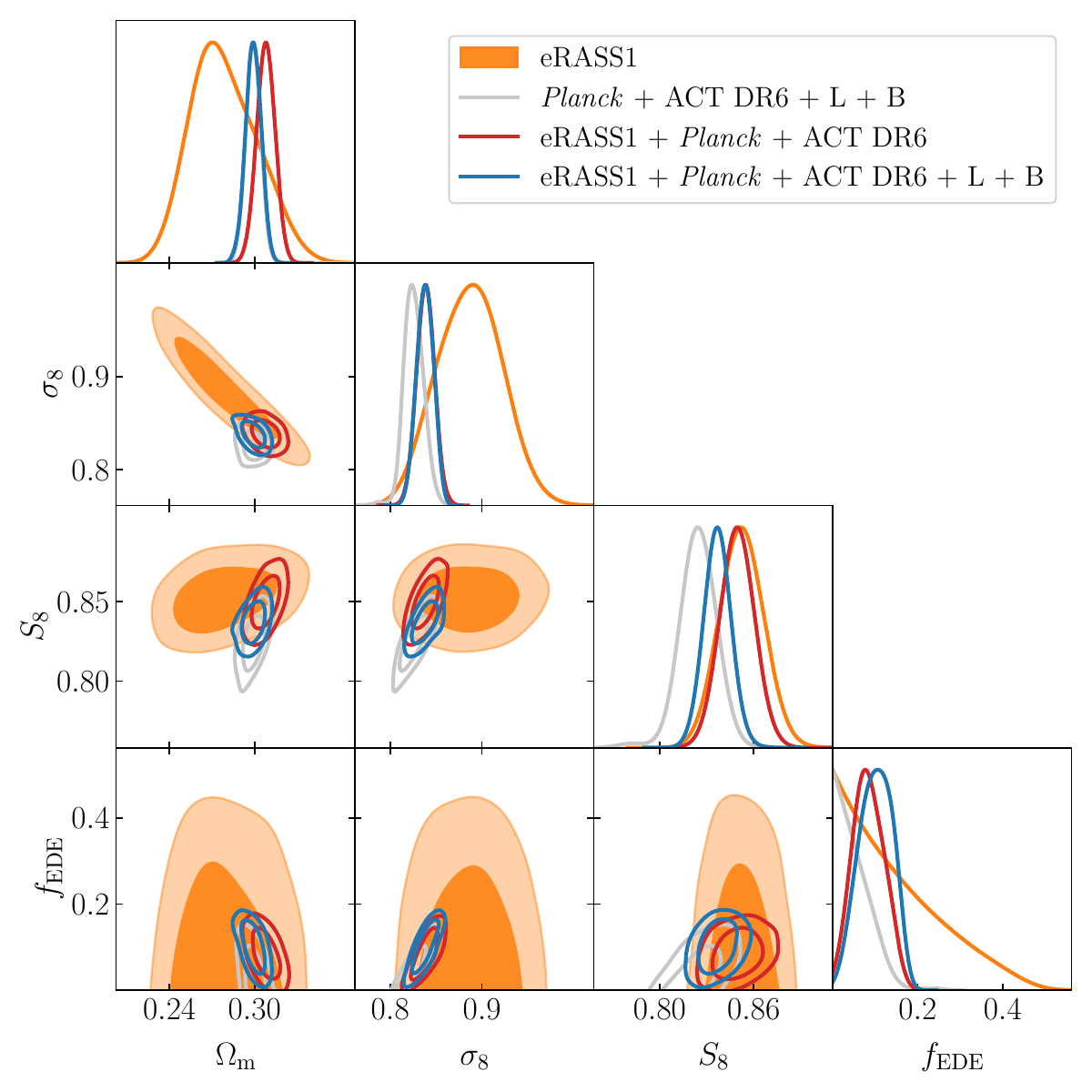}
    \caption{Marginalized posteriors for the parameters $\Omega_\mathrm{m}$, $\sigma_\mathrm{8}$, $S_8$ and $f_\mathrm{EDE}$ constrained for the eRASS1 EDE multiprobe joint analysis. In orange, we show the eRASS1-only constraints. The results of the eRASS1 joint analysis with \citetalias{PlanckCollaboration2020}, and \citetalias{Louis2025} are shown in solid blue lines. The reported results of the  \citetalias{PlanckCollaboration2020}, ACT DR6 CMB anisotropy and lensing, and BAO from DESI \citep{Adame2025} analysis are shown in light gray contours \citepalias{Calabrese2025}, and the results of this combination with eRASS1 are shown in solid red lines. When combined with CMB, BAO, and LSS data, eROSITA cluster number counts lead to a $3.2\sigma$ detection of the EDE fraction $f_\mathrm{EDE}=0.10^{+0.04}_{-0.03}$ and Hubble parameter of $H_{0}=\HZeroeRASSPACTLB$~km~s$^{-1}$~Mpc$^{-1}$.}
    \label{fig:Fig6}
\end{figure*}

Our eRASS1-only results agree within $0.15\,(1.4\sigma)$ for $S_8$ and $0.13\,(1.5\sigma)$ for $\sigma_8$ and $\Om$ with the constraints reported in \citetalias{Calabrese2025} for the combination of \citetalias{PlanckCollaboration2020} and \citetalias{Louis2025} CMB measurements, CMB lensing from \planck \citep{PlanckCollaboration2020b} and ACT~DR6 \citep{Qu2024,Madhavacheril2024} with BAO data from DESI \citep{Adame2025}. Given the acceptable agreement, we extend our analysis further using these posteriors. Together with measurements of structure growth and the CMB angular power spectra, weak gravitational lensing of CMB photons by large-scale structure (CMB lensing) and BAO provide constraints on late-time expansion and structure growth, which we consider largely independent of the eRASS1 analysis, since the two probes sample distinct redshift ranges. Naturally, the addition of CMB lensing and BAO data yields the strongest constraints on the structure parameters, shown in Tab.~\ref{tab:all_constraints}. There is a $\sim1\sigma$ shift towards lower inferred values of $\Omega_\mathrm{m}$ and $S_8$ compared to the eRASS1 analysis combined with \planck and ACT, shown in Fig.~\ref{fig:Fig6}. This is discussed in App.~\ref{subsec:CMBLensingBAO_Discussion}, showing good agreement with \citet{Chaussidon2025} and \citet{Poulin2025}, which find that EDE models jointly fit to the CMB and BAO acoustic datasets prefer lower $\Om$ and higher $H_\mathrm{0}$.

\begin{figure}[htb]
\centering
    \includegraphics[width=\linewidth]{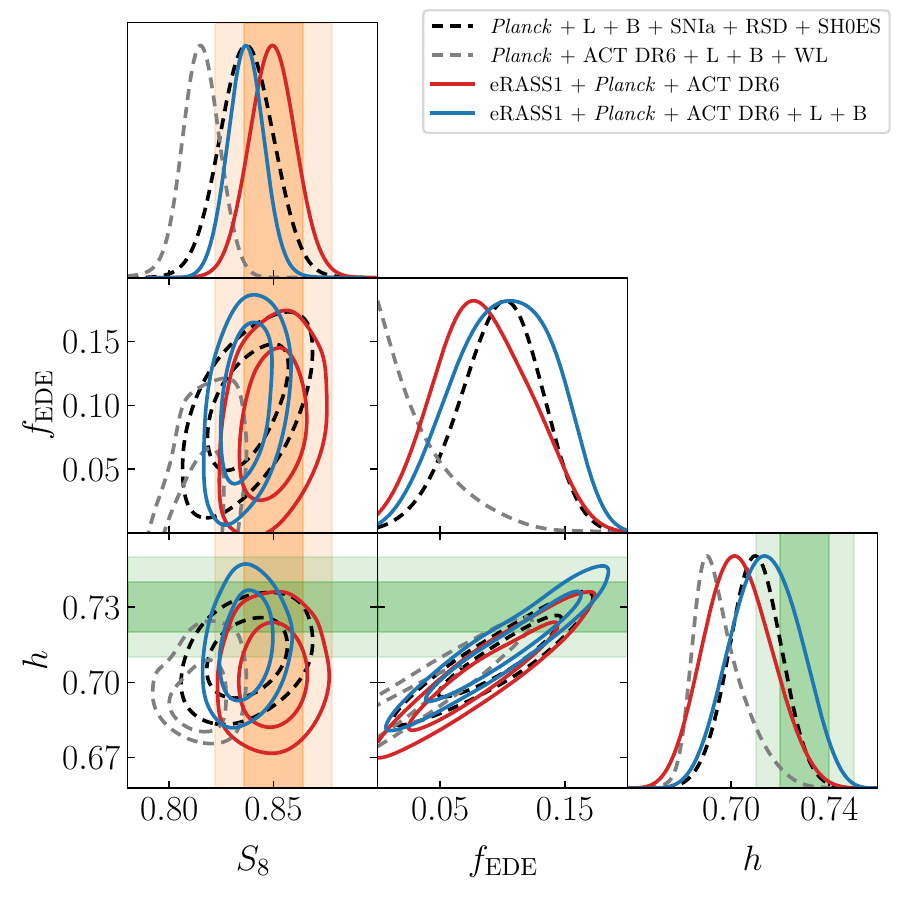}
    \caption{Marginalized posteriors showing the degeneracies between the parameters $S_8$, $f_\mathrm{EDE}$ and $h\equiv H_0/[100\,{\rm km}\,{\rm s}^{-1}\,{\rm Mpc}^{-1}]$ for the eRASS1 EDE multiprobe joint analysis. The eRASS1, \citetalias{PlanckCollaboration2020}, and \citetalias{Louis2025} joint analysis is shown in solid blue lines. The marginalized posteriors obtained for eRASS1, \citetalias{PlanckCollaboration2020}, ACT DR6 CMB anisotropy and lensing, and BAO from the DESI analysis \citep{Adame2025} are shown in solid red lines. In black dashed lines, we show the results reported in  \citet{Hill2020} using a baseline probe combination that includes \citetalias{Riess2022} data, and in gray dashed lines, we show the results reported in  \citetalias{Calabrese2025} when including WL data in the CMB and BAO analysis, through an $S_8$ prior from DES and KiDS. The vertical orange shaded area represents the $1\sigma$ and $2\sigma$ eRASS1-only constraints on $S_8$, and the green horizontal shaded area represents the $1\sigma$ and $2\sigma$ \citetalias{Riess2022} $H_0$ constraints. The discrepancy in the inferred value of $S_8$ between cosmic shear analyses and eRASS1 leads to differing constraints on $f_\mathrm{EDE}$ and $H_0$, due to the degeneracy of these parameters with $S_8$ present in the CMB data. Driven by the $S_8$ preference from eRASS1, our constraints are fully compatible with EDE multiprobe analyses that include \citetalias{Riess2022} SN~Ia data. }
    \label{fig:Fig7}
\end{figure}

\begin{figure}[htb]
\centering
    \includegraphics[width=\linewidth]{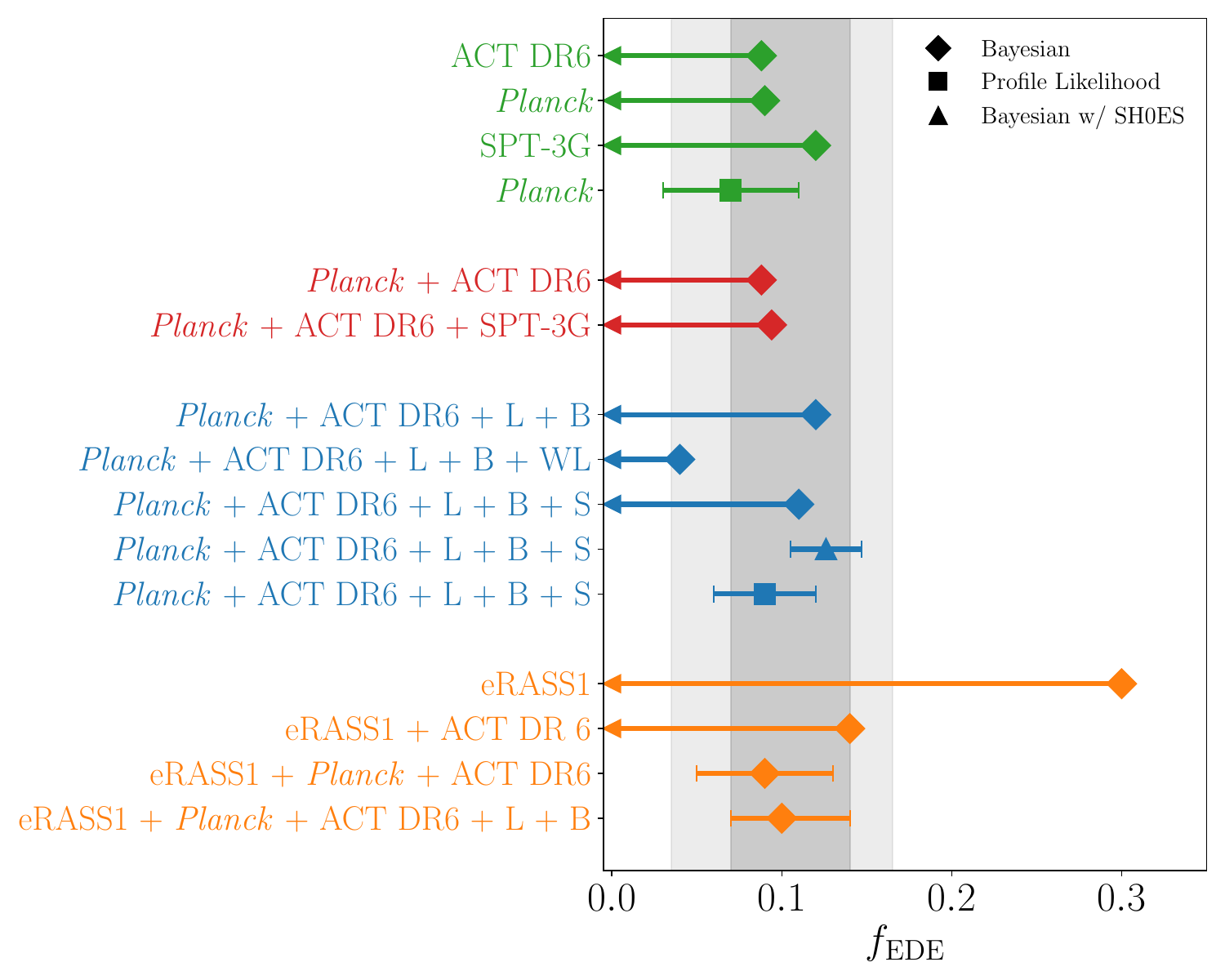}
    \caption{Representative whisker plot of the constraints obtained on $f_\mathrm{EDE}$, showing the differences in the constraints obtained using Bayesian analysis with and without a \citetalias{Riess2022} prior, and profile likelihood analyses. In green, we show the Bayesian constraints from single CMB probes for ACT~DR6 \citepalias{Calabrese2025},  \planck \citep{McDonough2024}, and SPT-3G \citep{Khalife2025}, as well as the \planck profile likelihood constraints \citep{Herold2023}. In red, we show CMB multiprobe constraints, for ACT and \planck \citepalias{Calabrese2025}, as well as the combination of SPT, ACT, and \planck \citep{Khalife2025}. In blue, we show a selected subsample of results from multiprobe EDE analyses building on top of ACT and \planck, including  CMB lensing (L) from ACT and \planck, DESI BAO, and DES WL \citepalias{Calabrese2025}, or Pantheon+ SN~Ia (S), from Bayesian and profile likelihood analyses \citep{Poulin2025}. Finally, the results from this work for a Bayesian analysis including eRASS1 cluster counts are shown in orange, with the gray-shaded areas representing our full probe combination $1\sigma$ and $2\sigma$ constraints on $f_\mathrm{EDE}$.}
    \label{fig:Fig8}
\end{figure}

The constraints obtained from this combination help break early-time probe degeneracies in the EDE parameter space, thereby enabling detection of EDE parameters. As summarized in Fig.~\ref{fig:Fig7} and Fig.~\ref{fig:Fig8}, the constraints represent our most significant $0.001\,(3.2\sigma)$ detection on $f_\mathrm{EDE}$ parameter of $\fEDEeRASSPACTLB$, and are fully consistent with previously reported non-zero EDE constraints from Bayesian analyses with cosmic distance calibration, as well as from profile-likelihood analyses. The inferred value of $H_\mathrm{0}=\HZeroeRASSPACTLB$~${\rm km}\,{\rm s}^{-1}\,{\rm Mpc}^{-1}$ is consistent within $1\sigma$ of the $H_\mathrm{0}=73.04\pm1.04$~${\rm km}\,{\rm s}^{-1}\,{\rm Mpc}^{-1}$ reported by \citetalias{Riess2022}, as well as the rest of the aforementioned late-time geometric probes.

As highlighted in the previous Sec.~\ref{subsec:CMB_combination}, while our structure growth parameter constraints are compatible within $2\sigma$ with joint Bayesian analyses of CMB and galaxy clustering data, the picture painted in our analysis by the preference for a higher $S_8$---and consequently the derived constraints on $f_\mathrm{EDE}$---is considerably different. In fact, when comparing our findings with multiprobe Bayesian analyses that further include SN~Ia data, the compatibility of the constraints on $\sigma_8$, $\Om$, $S_8$, and $f_\mathrm{EDE}$ becomes dependent on whether the SN~Ia data are distance-ladder calibrated. 

Bayesian analyses using eBOSS and BOSS with uncalibrated SN~Ia, \planck, and external BAO data constrain the cosmological parameters to $\Om=0.313\pm0.006$, $\sigma_8=0.814^{+0.007}_{-0.008}$ and $f_\mathrm{EDE}<0.0584$, reported in \citet{Gsponer2024}. While statistically compatible with our results, our constraints on $\sigma_8$ are $\sim2\sigma$ higher, leading to diverging interpretations for $f_\mathrm{EDE}$. In contrast, the same probe combination with the adoption of the \citetalias{Riess2022} calibration for SN~Ia
leads to a non-zero early dark energy fraction, $f_\mathrm{EDE}=0.118^{+0.025}_{-0.022}$, $\logzc= 3.621^{+0.17}_{-0.11}$, alongside shifts to $\Om=0.300\pm0.005$ and $\sigma_8=0.838\pm0.010$ \citep{Gsponer2024}. Notably, these shifts, induced by the inclusion of cosmic distance ladder data, improve agreement with both our eRASS1-only constraints and our joint analysis with external probes to within $\sim1\sigma$. Similar results are found in other studies, where combining \citetalias{Riess2022}-calibrated SN~Ia with CMB, BAO, and RSD data yields $f_\mathrm{EDE}=0.107\pm0.023$, $\logzc=3.58^{+0.05}_{-0.15}$, $\Om=0.302 \pm 0.005$ and $S_8=0.838\pm0.012$ \citep{Efstathiou2023}, with minor variations depending on the choice of CMB likelihood and the inclusion of CMB lensing \citep{Smith2020, Poulin2025}. 

As qualitatively illustrated in Fig.~\ref{fig:Fig8} and detailed in App.~\ref{subsec:gof_multiprobe}, the eRASS1 preference for a mildly higher $S_8$ appears to act as a data-driven anchor---analogous to the preference for a higher $H_0$ when including calibrated SN~Ia---which helps mitigate the $f_\mathrm{EDE}\sim 0$ prior volume bias and recovers a hint of detection in our multiprobe constraints. Consequently, our results align with frequentist analyses that avoid these volume effects, such as the reported profile likelihood constraints on $f_\mathrm{EDE}=0.09^{+0.030}_{-0.033}$, $H_\mathrm{0}=71.96^{+1.08}_{-0.99}$~${\rm km}\,{\rm s}^{-1}\,{\rm Mpc}^{-1}$ from a dataset combining \planck and ACT anisotropy and lensing, BAO from DESI, and Pantheon+ SN~Ia \citep{Poulin2025}. 

Cosmic shear analyses also constrain $S_8$ values to be $2-3\sigma$ lower than those inferred from eRASS1 or CMB data, and their inclusion in EDE multiprobe analyses is used to argue that LSS data disfavor EDE as a resolution to the Hubble tension.
The EDE probe combination analysis reported in \citet{Hill2020} serves as a particularly apt reference for illustrating the conflicting EDE implications between our analysis and the constraints obtained when including cosmic shear. Our multiprobe results are fully consistent within $1\sigma$ with the baseline joint analysis including CMB, lensing, BAO, RSD and \citetalias{Riess2019} used in \citet{Hill2020} to report the structure parameter constraints $S_8=0.837\pm0.013$, $\sigma_8=0.833 \pm 0.010 $ and $\Om=0.3025\pm0.0051$, as well as the expansion rate $H_\mathrm{0}=70.98\pm1.05$~${\rm km}\,{\rm s}^{-1}\,{\rm Mpc}^{-1}$ and the EDE parameters $f_\mathrm{EDE}=0.098\pm0.032$, $\log (z_\mathrm{c})=3.63^{+0.17}_{-0.10}$, $\theta_\mathrm{i}=2.58^{+0.29}_{-0.09}$. However, when cosmic shear measurements are incorporated (using DES-Y1 3x2pt data with $S_8 = 0.773^{+0.026}_{-0.020}$, HSC with $S_8=0.780^{+0.030}_{-0.033}$ and KV-450 with $S_8=0.737^{+0.040}_{-0.036}$), the preference for lower $S_8 = 0.809 \pm 0.010$ drives $f_\mathrm{EDE}<0.112$ to be consistent with zero via the $S_8$ and $f_\mathrm{EDE}$ parameter degeneracy. This is further underscored by the inclusion of cosmic shear data in the probe combination without \citetalias{Riess2019}, reported to decrease the preference for EDE, with $S_8=0.807 \pm 0.009$, $f_\mathrm{EDE}<0.06$ and $H_\mathrm{0}=68.92^{+0.57}_{-0.59}$~${\rm km}\,{\rm s}^{-1}\,{\rm Mpc}^{-1}$. These stringent constraints on $f_\mathrm{EDE}$ are replicated in \citet{McDonough2024}, and in the \citetalias{Calabrese2025} probe combination analysis that includes $S_8$ measurements from DES-Y3 and KiDS-1000 surveys, which significantly tightens the upper limit on EDE \citep{Calabrese2025}, as shown in gray in Fig.~\ref{fig:Fig7}. Because massive cluster abundances primarily probe the linear-structure growth regime, whereas cosmic shear also probes smaller scales affected by non-linear physics, this $S_8$ discrepancy may stem from fundamentally different physical sensitivities. We explore the origins of this discrepancy, the impact of recent weak lensing calibrations, and the broader physical implications of our EDE constraints in Section \ref{sec:discussion}.

\section{Discussion}
\label{sec:discussion}

Ultimately, the $2-3\sigma$ discrepancy in $S_8$ between cosmic shear analyses and the eRASS1 cluster abundances directly mirrors the conflicting constraints placed on $f_\mathrm{EDE}$ in joint EDE analyses with early-Universe data. To determine whether LSS data truly disfavor or support the viability of EDE as a resolution to the Hubble tension, it is therefore relevant to discuss the potential origins of the discrepancy.

First, we address potential systematic differences between cosmic shear and cluster-count analyses that use the same lensing data. Extracting both the cosmic shear and cluster lensing signals from the same optical catalogs introduces shared systematics in galaxy shapes and photometric redshifts (photo-$z$), but the propagation of these uncertainties differs fundamentally.  Cluster mass calibration relies on localized, pointed tangential shear profiles around identified cluster centers, meaning photo-$z$ errors and multiplicative shear biases scale linearly. Conversely, cosmic shear measures the shear autocorrelation across the continuous field and is thus impacted by these uncertainties quadratically.
Reflecting this sensitivity, recent cosmic shear \lcdm results from KiDS and HSC with updated photo-$z$ distribution estimation and calibration report markedly higher values of $S_8$ \citep{Wright2025, Choppin2025}. 

Furthermore, as discussed in \citetalias{Ghirardini2024}, while both LSS analyses probe the same matter density field, they are sensitive to structure growth on distinct mass and angular scales and, consequently, are subject to different modeling considerations. 
Cluster abundances constrain the high-mass tail of the halo mass function, sensitive to massive galaxy clusters with $M\gtrsim10^{14}h^{-1}M_\odot$. These collapsed massive halos preserve information from the linear fluctuation regime at the time their growth detached from the background expansion. Overall, cluster abundances are subject to systematics arising from the theoretical foundation of cluster number counts, as well as observational biases from cluster detection, selection effects, and mass calibration. Within the \lcdm framework, these are accounted for in our analysis. At the statistical precision of the eRASS1 data, and given that EDE behaves like \lcdm at late times, the EDE model is not expected to introduce additional significant systematics (as discussed in Sec.~\ref{sec:validation} and App.~\ref{sec:HMF_Systematics}).

On the other hand, while the cosmic shear signal at very large scales can be predicted with remarkable accuracy, predictions at intermediate and smaller scales are sensitive to several potential astrophysical systematics, such as the non-linear power spectrum calculation or the intrinsic alignment (IA) of galaxy shapes with the lensing potential \citep[see e.g.,][for detailed discussions]{Mandelbaum2018,Huterer2023}. 
Cluster abundances, because of their sensitivity to the linear regime, are less affected by the former, and restricting our analysis to sufficiently small scales where the cluster's potential dominates the lensing signal ensures that our results are unaffected by the latter. Notably, a cosmic shear analysis with a reduced blue galaxy sample of DES Y3, argued to be less impacted by IA, reports higher inferred values of $S_8=0.822^{+0.019}
_{-0.020}$ \citep{McCullough2026}. 

Alternatively, discrepancies in the inferred values of $S_8$ could potentially arise from scale-dependent effects on the matter power spectrum. Cluster counts and cosmic shear are two fundamentally different statistics of the matter power spectrum that compress information across different scales into the inherently linear $S_8$ parameter. Because of this, even in a self-consistent cosmology, scale-dependent suppression can lead to significant differences when projected onto $S_8$. 

Baryon feedback, suppressing structure growth on galactic and group scales, provides a natural astrophysical mechanism. For instance, \citet{Amon2022} highlighted that applying a more energetic feedback model to KiDS-1000 data can increase $S_8$ to CMB-compatible levels. More energetic feedback models could be well motivated by recent X-ray and kinetic Sunyaev-Zeldovich \citep[kSZ effect,][]{SunyaevZeldovich1972} observations, which have revealed diffuse gas beyond the ranges predicted by N-body simulations \citep{Zhang2026, Hadzhiyska2025}. Cosmic shear analyses, which cross-correlate with these probes to inform baryon feedback modeling, yield higher values of $S_8$ \citep{Arico2023, Bigwood2025}.

Scale-dependent suppression could also stem from departures from the standard cosmological model, such as dark matter interactions or primordial non-Gaussianities \citep[e.g., ][]{Poulin2023b, Stahl2024, Teixeira2025}. The combination of these mechanisms with the EDE expansion boost could simultaneously resolve the discrepant cosmological observations of $H_\mathrm{0}$ and $S_8$, at the expense of additional parameters \citep{Stahl2025, Simon2025}. 

In view of the underlying systematics and the distinct modeling choices used to address them across different analyses, recent cosmic shear constraints on $S_8$ exhibit significant scatter. The latest KiDS-legacy analysis, featuring updated photo-$z$ distribution estimation and calibration, together with expanded baryon feedback models, reports values of $S_8=0.815^{+0.015}_{-0.021}$ compatible with CMB constraints \citep{Wright2025}. The recent Unions-3500 analyses also report a larger, albeit less constraining, $S_8=0.83^{+0.07}_{-0.08}$ \citep{Goh2026}. In contrast, the DES Y6 3x2pt analysis still reports a lower $S_8=0.789\pm0.012$ \citep{Abbott2026}. Although this differs by $2.6\sigma$ from combined CMB constraints, the full-space parameter difference is a less significant $1.8\sigma$. Importantly, this analysis is performed with aggressive scale cuts, indicating that the discrepancy might persist on the linear scale.

Notwithstanding the scatter in the constraints on $S_8$, even CMB-compatible cosmic shear results remain $\gtrsim1.5\sigma$ below the eRASS1 cluster abundance constraints. Cluster abundance analyses also show internal scatter within \lcdm \citepalias[see e.g., ][for a more detailed comparison]{Ghirardini2024}. These analyses differ from eRASS1 in the cluster selection methodology and the associated modeling. SPT-selected clusters yield $S_8=0.795\pm0.029$, values compatible with CMB and eRASS1 results, but respectively $\sim1.1\sigma$ and $\sim1.8\sigma$ lower \citep{Bocquet2024}. Optically selected clusters from DES Y3, on the other hand, also report enhanced values of $S_8=0.864\pm0.035$ \citep{Abbott2025}, fully compatible with eRASS1 constraints. Given that these independent cluster constraints remain broadly compatible, the mildly elevated $S_8$ derived from eRASS1 may represent a statistical upscatter---naturally expected and potentially driven by differing cluster selection effects---within values compatible with CMB-inferred \lcdm constraints. Future eRASS surveys and analysis efforts will increase the precision of cosmological constraints, enabling a determination of whether the preference for a slightly larger $S_8$ reflects a statistical fluctuation or an indication of enhanced growth relative to \lcdm CMB predictions. In the EDE cosmological context, as future analyses increase in precision, greater attention to modeling choices will be required. Particularly, further EDE N-body simulations with a broader exploration of cosmological realizations will be key to either calibrate a self-consistent $f(\sigma)$ in EDE cosmologies, or validate the approach taken in this work at higher precision. 

Interestingly, the mild preference found in the eRASS1 analysis could be consistent with hints of enhanced early structure growth from high-$z$ JWST observations, despite probing different epochs and scales of structure formation. The unexpectedly large populations of collapsed halos observed---in terms of an elevated abundance of spectroscopically-confirmed bright ultraviolet galaxies at $z\gtrsim10$ \citep[e.g.,][]{Curtis-Lake2023, Harikane2024}, as well as several extremely massive galaxy candidates across a
broader redshift range $z\sim3-12$ \citep[e.g.,][]{Labbe2023, Glazebrook2024}---could present an independent challenge to the standard \lcdm structure formation paradigm. Although explanations remain degenerate with modeling choices for the galaxy–halo connection at high redshifts \citep{Cochrane2025}, the most massive galaxy candidates might lie at the limit of the \lcdm halo mass function \citep{Boylan-Kolchin2023, Lovell2023}. EDE cosmologies feature increased power on small scales, driven by shifts towards $n_s\sim1$ required to fit the Silk damping scale from CMB data \citep{Jiang2022}. The consequent accelerated halo formation leads to an overall enhancement of early structure growth (as can also schematically be seen in Fig.~\ref{fig:Fig3}, where the predicted late-time structure growth differences are mild, but at higher redshifts the EDE model fitted to the CMB naturally leads to an increasing abundance of massive halos), offering a natural resolution to the JWST discrepant observations \citep{Shen2024,Shen2025}.

Combined with other findings on the ability of EDE to provide unified explanations of the Hubble tension and the observed cosmic birefringence \citep{Kochappan2025}, as well as modestly improving the agreement between the acoustic measurements of DESI BAO and CMB data \citep{Khalife2025,Chaussidon2025}, our results highlight EDE as a promising extension of \lcdm to explain cosmological observables where the current paradigm seems to fall short. Consequently, we conclude that based on the joint analysis of eRASS1 cluster abundances with early-Universe and cosmic-shear-independent LSS data, EDE remains a promising solution to the Hubble tension while yielding structure growth constraints compatible with LSS analyses. 


\section{Conclusions}

This work presents the first constraints on the cosmological parameters for a flat \lcdm cosmology with an additional Early Dark Energy (EDE) component, parametrized by the $f_\mathrm{EDE},\log_{10}z_\mathrm{c}$ and $\theta_\mathrm{i}$, through the cluster mass function. The volume-limited catalog used in this work includes 4196 clusters of galaxies detected in the first eROSITA All-Sky Survey in the Western Galactic Hemisphere, with redshifts ranging from 0.1 to 0.45 \citep[see the cosmology sample in][]{Bulbul2024, Kluge2024}. While the sample purity is above 95\% \citep{Kluge2024}, the remaining contamination is identified and removed by the mixture model. The weak lensing mass calibration is derived from optical surveys, including DES, KiDS, and HSC, which have substantial overlap with eROSITA \citep{Grandis2024a, Kleinebreil2024, Okabe2025}. The selection function, which models the survey's detection probability and completeness, is derived from twin mock observations of eROSITA and folded into our analysis \citep{Clerc2024, Seppi2022}. 

Incorporating early dark energy models into the flat $\Lambda$CDM cosmology framework presented in \citet{Ghirardini2024} requires modifications to the standard halo mass function of \citet{Tinker2008}. Accordingly, we adjust the matter power spectrum using the \texttt{EarlyQuintessence} module \citep{Smith2020} implemented in \texttt{CAMB} \citep{Lewis2000}. To validate our implementation, we compare the resulting theoretical predictions with N-body simulations that include an early dark energy component \citep{Klypin2021}. In our inference pipeline, the eROSITA cluster number counts are modeled within our framework, which includes the EDE component, and are fit using a forward-modeled Bayesian likelihood.

The full Bayesian analysis of the eRASS1 cluster number counts yields an upper limit on the fraction of early dark energy $f_\mathrm{EDE}<\fEDEeRASS$. The inferred values of the key cosmological parameters, the normalization of matter density fluctuations $\sigma_{8}$, the total matter density parameter $\Om$, and the $ S_{8}$ parameter, are consistent within $1\sigma$ with the standard \lcdm model constraints reported by \citetalias{Ghirardini2024}. This agreement demonstrates the robustness of our analysis. 
Our constraints on the key cosmological parameters are compatible within $1.7\sigma$ of those reported in the literature using CMB data.

This work provides the first use of the cluster halo mass function in the EDE framework, demonstrating that its combination with CMB data can recover $f_\mathrm{EDE}$ and Hubble constant measurements consistent with those inferred from the cosmic distance ladder. To break parameter degeneracies, we combine and compare our results with posterior constraints from complementary, orthogonal probes---provided their $\sigma_8$ and $\Om$ constraints agree with the eROSITA results at the $2\sigma$ confidence level. These comprise CMB temperature and polarization anisotropy measurements from ACT~DR6 and \planck \citep{Calabrese2025, PlanckCollaboration2020}, CMB lensing measurements \citep{Madhavacheril2024,Qu2024,PlanckCollaboration2020b}, and BAO measurements from DESI DR1 \citep{Adame2025, Qu2024b}. While the combination of the eRASS1 cluster number counts with ACT CMB anisotropies yields an upper limit of $f_\mathrm{EDE}<0.14$, the joint analysis with \planck and ACT combined posteriors yields a $2.8\sigma$ hint of EDE ($f_\mathrm{EDE}=\fEDEeRASSPACT$). This represents the first emergence of such a signal from a probe sensitive to the growth of structure and the CMB, albeit at marginal significance. Due to the degeneracy between $H_\mathrm{0}$ and $f_\mathrm{EDE}$ introduced by the early-Universe data, this joint analysis also yields a Hubble parameter measurement of $H_{0}=\HZeroeRASSPACT$~${\rm km}\,{\rm s}^{-1}\,{\rm Mpc}^{-1}$, which eROSITA can only measure in combination with other probes. Further including CMB lensing and BAO strengthens constraints on late-time expansion and structure growth, and reveals a $3.2\sigma$ detection of $f_\mathrm{EDE}=\fEDEeRASSPACTLB$. The resulting Hubble parameter, $H_\mathrm{0}=\HZeroeRASSPACTLB$~${\rm km}\,{\rm s}^{-1}\,{\rm Mpc}^{-1}$, is in full agreement with the constraints reported using \citetalias{Riess2022} SN~Ia measurements.

Our eRASS1 analysis, as well as its subsequent combination with CMB data, yields higher values of the parameter $\sigma_8$ (and $S_{8}$) than those from other LSS probes, in particular cosmic shear. The agreement with analyses that include BAO and galaxy clustering remains within $2\sigma$ and is notably improved to $\sim1\sigma$ when calibrated SN~Ia data are included, yielding non-zero constraints on EDE. The tension with cosmic shear measurements is on the order of $\sim3\sigma$, and is not improved with the addition of SN~Ia data. This preference for higher-structure growth normalization drives our higher inferred $f_\mathrm{EDE}$ and $H_\mathrm{0}$. The potential reasons for the discrepancy are discussed, drawing upon \citetalias{Ghirardini2024}, and relevant cosmic shear cosmology work \citep[e.g.,][]{Arico2023, Bigwood2025, Wright2025}.

We have shown that, within the redshift range probed by eRASS1, our modified EDE HMF reproduces the cluster-count population from the \citet{Klypin2021} EDE simulations to within the margins required for precision cosmology. Because our baseline HMF is calibrated in \lcdm to relate non-linear structure growth to the variance of the underlying linear density field, our approach assumes that the impact of EDE---which redshifts rapidly after recombination---is primarily driven by modifications to the linear matter power spectrum. While our comparison with simulations validates this assumption, it remains limited by the current lack of large-scale structure simulations that cover a broader range of EDE parameter realizations. A natural next step would be to derive the halo mass function directly from such N-body simulations and apply it to probes of the growth of structure. However, such a calibration is not yet available in the literature.

On the observational side, eROSITA has already completed more than 4 all-sky surveys (eRASS5), providing data that will enable statistical precision in constraints on fundamental cosmological parameters at a level comparable to CMB probes. When combined with halo mass functions derived from EDE simulations, these data will help mitigate systematics in the mass–observable relation and may enable robust tests of EDE models. The tentative preference for an EDE signal, emerging from combining eRASS1 cluster abundances with CMB and other probes, underscores the need for further simulations and modeling to reduce systematic uncertainties in the halo mass function that describes structure growth in EDE cosmologies. This will be crucial for confirming the viability of the EDE scenario and for assessing whether current large-scale structure data disfavor or support EDE as a potential resolution to the Hubble tension.

\begin{acknowledgement}

The authors thank the referee for their helpful and constructive comments on the draft.

This work is based on data from eROSITA, the soft X-ray instrument aboard SRG, a joint Russian-German science mission supported by the Russian Space Agency (Roskosmos), in the interests of the Russian Academy of Sciences represented by its Space Research Institute (IKI), and the Deutsches Zentrum f{\"{u}}r Luft und Raumfahrt (DLR). The SRG spacecraft was built by the Lavochkin Association (NPOL) and its subcontractors and is operated by NPOL with support from the Max Planck Institute for Extraterrestrial Physics (MPE).

The development and construction of the eROSITA X-ray instrument was led by MPE, with contributions from the Dr. Karl Remeis Observatory Bamberg \& ECAP (FAU Erlangen-Nuernberg), the University of Hamburg Observatory, the Leibniz Institute for Astrophysics Potsdam (AIP), and the Institute for Astronomy and Astrophysics of the University of T{\"{u}}bingen, with the support of DLR and the Max Planck Society. The Argelander Institute for Astronomy of the University of Bonn and the Ludwig Maximilians Universit{\"{a}}t Munich also participated in the science preparation for eROSITA.

The eROSITA data shown here were processed using the \esass software system developed by the German eROSITA consortium.
\\

E. Bulbul, E. Artis, V. Ghirardini, A. Liu, S. Zelmer, and X. Zhang acknowledge financial support from the European Research Council (ERC) Consolidator Grant under the European Union’s Horizon 2020 research and innovation program (grant agreement CoG DarkQuest No 101002585). N. Clerc was financially supported by CNES. T. Schrabback and F. Kleinebreil acknowledge support from the German Federal
Ministry for Economic Affairs and Energy (BMWi) provided
through DLR under projects 50OR2002, 50OR2106, and 50OR2302, as well as the support provided by the Deutsche Forschungsgemeinschaft (DFG, German Research Foundation) under grant 415537506.

\\

The Innsbruck authors acknowledge the support provided by
The Austrian Research Promotion Agency (FFG) and the Federal Ministry of the Republic of Austria for Climate Action, Environment, Energy, Mobility,
Innovation and Technology (BMK) via the Austrian Space Applications Programme with grant numbers 899537, 900565, and 911971.

\\

Funding for the DES Projects has been provided by the U.S. Department of Energy, the U.S. National Science Foundation, the Ministry of Science and Education of Spain, the Science and Technology FacilitiesCouncil of the United Kingdom, the Higher Education Funding Council for England, the National Center for Supercomputing Applications at the University of Illinois at Urbana-Champaign, the Kavli Institute of Cosmological Physics at the University of Chicago, the Center for Cosmology and Astro-Particle Physics at the Ohio State University, the Mitchell Institute for Fundamental Physics and Astronomy at Texas A\&M University, Financiadora de Estudos e Projetos, Funda{\c c}{\~a}o Carlos Chagas Filho de Amparo {\`a} Pesquisa do Estado do Rio de Janeiro, Conselho Nacional de Desenvolvimento Cient{\'i}fico e Tecnol{\'o}gico and the Minist{\'e}rio da Ci{\^e}ncia, Tecnologia e Inova{\c c}{\~a}o, the Deutsche Forschungsgemeinschaft, and the Collaborating Institutions in the Dark Energy Survey.

The Collaborating Institutions are Argonne National Laboratory, the University of California at Santa Cruz, the University of Cambridge, Centro de Investigaciones Energ{\'e}ticas, Medioambientales y Tecnol{\'o}gicas-Madrid, the University of Chicago, University College London, the DES-Brazil Consortium, the University of Edinburgh, the Eidgen{\"o}ssische Technische Hochschule (ETH) Z{\"u}rich,  Fermi National Accelerator Laboratory, the University of Illinois at Urbana-Champaign, the Institut de Ci{\`e}ncies de l'Espai (IEEC/CSIC), the Institut de F{\'i}sica d'Altes Energies, Lawrence Berkeley National Laboratory, the Ludwig-Maximilians Universit{\"a}t M{\"u}nchen and the associated Excellence Cluster Universe, the University of Michigan, the National Optical Astronomy Observatory, the University of Nottingham, The Ohio State University, the OzDES Membership Consortium, the University of Pennsylvania, the University of Portsmouth, SLAC National Accelerator Laboratory, Stanford University, the University of Sussex, and Texas A\&M University.

Based on observations made with ESO Telescopes at the La Silla Paranal Observatory under program IDs 177.A-3016, 177.A-3017, 177.A-3018, and 179.A-2004, and on data products produced by the KiDS consortium. The KiDS production team acknowledges support from: Deutsche Forschungsgemeinschaft, ERC, NOVA, and NWO-M grants; Target; the University of Padova, and the University Federico II (Naples).

The HSC
instrumentation and software were developed by the National Astronomical Ob-
servatory of Japan (NAOJ), the Kavli Institute for the Physics and Mathematics
of the Universe (Kavli IPMU), the University of Tokyo, the High Energy Accelerator Research Organization (KEK), the Academia Sinica Institute for Astron-
omy and Astrophysics in Taiwan (ASIAA), and Princeton University. Funding
was contributed by the FIRST program from the Japanese Cabinet Office, the
Ministry of Education, Culture, Sports, Science and Technology (MEXT), the
Japan Society for the Promotion of Science (JSPS), Japan Science and Technology Agency (JST), the Toray Science Foundation, NAOJ, Kavli IPMU, KEK,
ASIAA, and Princeton University.

This paper utilizes software developed for the Large Synoptic Survey Telescope. We thank the LSST Project for making their code available as free software at  http://dm.lsst.org

\\

This work made use of the following Python software packages: 
SciPy\footnote{https://scipy.org/} \citep{Virtanen2020SciPy}, 
Matplotlib\footnote{https://matplotlib.org/} \citep{Hunter2007matplotlib}, 
Astropy\footnote{https://www.astropy.org/} \citep{Astropy2022}, 
NumPy\footnote{https://numpy.org/} \citep{Harris2020},
CAMB \citep{Lewis2000},
pyCCL\footnote{https://github.com/LSSTDESC/CCL} \citep{Chisari2019},
GPy\footnote{https://github.com/SheffieldML/GPy} \citep{gpy2014},
climin\footnote{https://github.com/BRML/climin} \citep{Bayer2015},
ultranest\footnote{https://github.com/JohannesBuchner/UltraNest/} \citep{Buchner2021}
emcee\footnote{https://emcee.readthedocs.io/} \citep{emcee2013}, Colossus\footnote{https://bdiemer.bitbucket.io/colossus/} \citep{Diemer2018}

\end{acknowledgement}

\bibliography{references.bib}
\begin{appendix}
\section{Impact of \texorpdfstring{$H_\mathrm{0}$}{H0} priors on eRASS1 analysis}
\label{sec:H0_priors}
 Cluster counts are largely insensitive to $H_\mathrm{0}$ due to its degeneracy with the other cosmological parameters, and thus cannot constrain it to precisions comparable with geometric probes \citep{Ghirardini2024}. We test the effects of using different priors for $H_\mathrm{0}$ on our constraints in EDE. These include Gaussian priors from the \citetalias{Riess2022} analysis with $\sim\mathcal{N}(73.0, 1.0)$, a prior from the \citetalias{PlanckCollaboration2020} \texttt{NPIPE} analysis with $\sim\mathcal{N}(68.1, 0.8)$ \citep{Efstathiou2023}, and the wider Gaussian prior with $H_\mathrm{0}\sim\mathcal{N}(70, 5)$.

\begin{table}[htbp]
\caption{eRASS1-only marginalized constraints on the relevant EDE cosmological parameters under different $H_\mathrm{0}$ priors.}
\centering
\begin{tabular}{lccc}
\hline
\hline
Parameter & $\sim\mathcal{N}(68.1,0.8)$ & $\sim\mathcal{N}(73,1)$ & $\sim\mathcal{N}(70,5)$ \\
\hline
$\Omega_{\mathrm{m}}$ & $0.27\pm0.02$ & $0.28\pm0.02$ & $0.27\pm0.02$ \\
$\sigma_{8}$ & $0.88\pm0.03$ & $0.88\pm0.03$ & $0.89\pm0.03$  \\
$S_{8}$ & $0.84\pm0.01$& $0.85\pm0.01$ & $0.85\pm0.01$  \\
$f_{\mathrm{EDE}}$ & $<0.3$ & $<0.3$ & $<0.30$ \\
$\log_{10}z_\mathrm{c}$ & $3.2^{+0.4}_{-0.5}$ & $3.4^{+0.4}_{-0.6}$ & $3.3^{+0.2}_{-0.7}$ \\
\hline
\hline
\end{tabular}
\tablefoot{Uncertainties represent 68\% marginalized constraints, except for the upper limit constraints on $f_{\mathrm{EDE}}$ given at a 95\% confidence interval.}
\label{tab:h0_priors}
\end{table}
\begin{figure}[htb]
\centering
    \includegraphics[width=\linewidth]{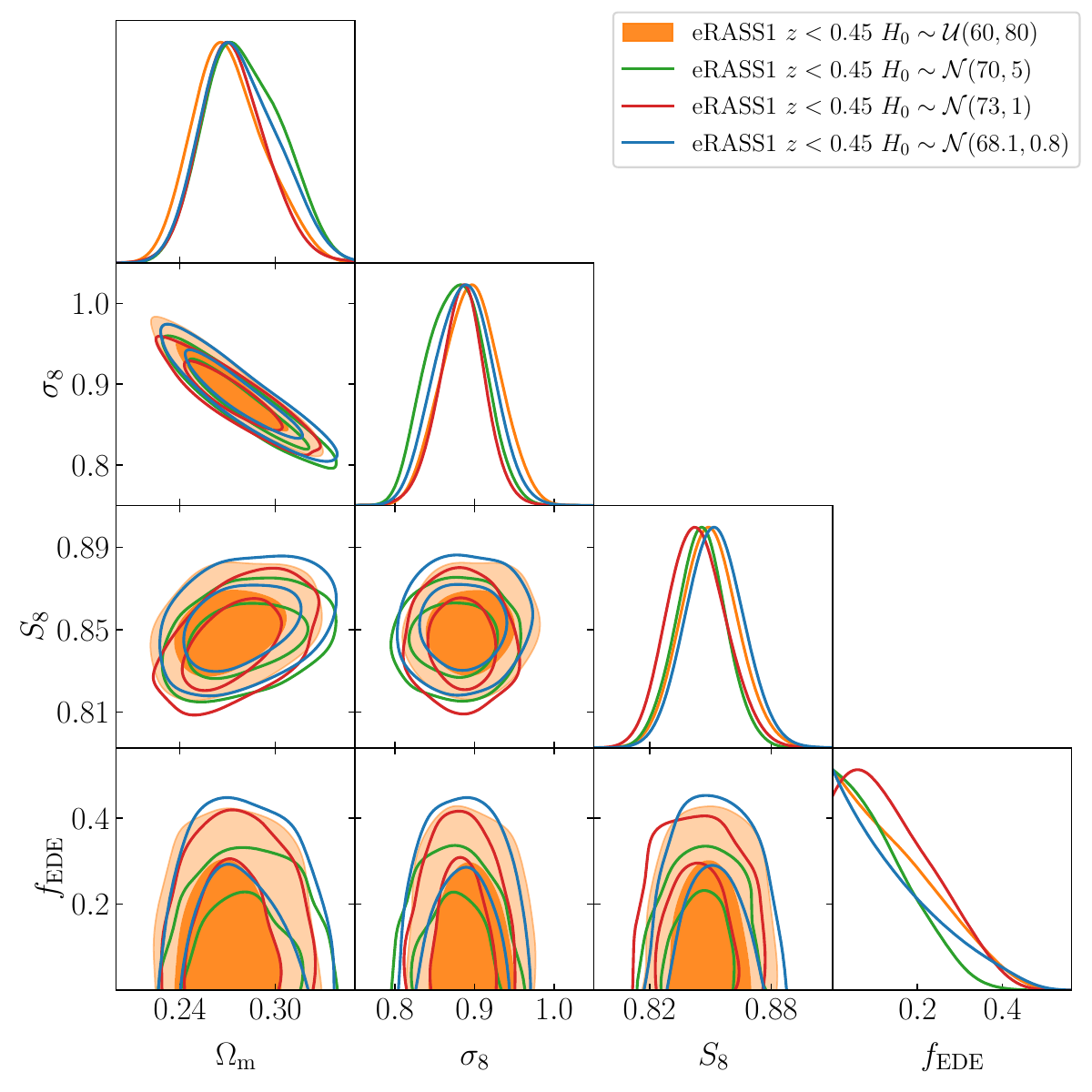}
    \caption{Robustness of the EDE constraints obtained for eRASS1 $0.1<z<0.45$ against choice of prior on $H_\mathrm{0}$: Contour plots for $\Om$, $\sigma_8$, $S_8$ and $f_\mathrm{EDE}$, with the main results using $H_\mathrm{0}\sim\mathcal{U}(60,80)$ shown in orange. The constraints remain consistent against this prior choice, with compatibility with early-Universe probes only marginally affected.}
    \label{fig:z045_eRASS1_H0_priors}
\end{figure} 

 As shown in Table \ref{tab:h0_priors} and Fig.~\ref{fig:z045_eRASS1_H0_priors},  the $\Omega_\mathrm{m}$ and $\sigma_\mathrm{8}$ constraints remain in full agreement when changing the $H_\mathrm{0}$ prior. As such, we find that the agreement with early-time probe EDE analyses for $S_8$ and the $\sigma_\mathrm{8}$ and $\Omega_\mathrm{m}$ plane does not depend on the $H_\mathrm{0}$ prior used. Our upper limits on $f_{\mathrm{EDE}}$ remain consistent regardless of assumptions regarding the $H_0$ prior from discrepant geometric probes. This underscores that cluster counts are primarily sensitive to the EDE-induced structure growth suppression, rather than to the background expansion rate $H_0$.

\section{Tests on \texorpdfstring{$f(\sigma)$}{multiplicity function}}
\label{sec:HMF_Systematics}

Within the context of precision cosmology with cluster abundances, the choice of multiplicity function $f(\sigma)$---which relates the number density of halos to the variance of the underlying linear matter density field---can have a non-negligible impact on the constraints obtained. For instance, in the \citetalias{Ghirardini2024} \lcdm analysis, changes to the HMF through the use of a different $f(\sigma)$ calibration introduce shifts within $1\sigma$ in the constraints obtained on $\sigma_8$ and $\Om$.

We seek to investigate the robustness of our EDE analysis against such theoretical systematics by performing two distinct tests. First, we replace our baseline HMF calibration from \citet{Tinker2008} with the alternative fitting function provided by \citet{Despali2016}. Second, we test our constraints when marginalizing over the theoretical uncertainty of the HMF. Following the formalism introduced by \citet{Costanzi2019}, we add two nuisance parameters that modify the normalization and slope of the \citet{Tinker2008} HMF:
\begin{equation}
    \frac{\dd n}{\dd\ln M} = \left(\frac{\dd n}{\dd\ln M}\right)_\mathrm{Tinker08} \left[s\ln\left(\frac{M}{M_*}\right) + q\right].
\end{equation}
To account for a systematic budget of $\sim 5\%$, we adopt Gaussian priors for the normalization parameter $q \sim \mathcal{N}(1, 0.035)$ and the slope parameter $s \sim \mathcal{N}(0, 0.02)$, with the pivot mass set to $\log(M_*) = 13.8\, h^{-1}M_\odot$, as done in similar analyses \citep{Ghirardini2024, Bocquet2024}.

\begin{figure}[htb]
\centering
    \includegraphics[width=\linewidth]{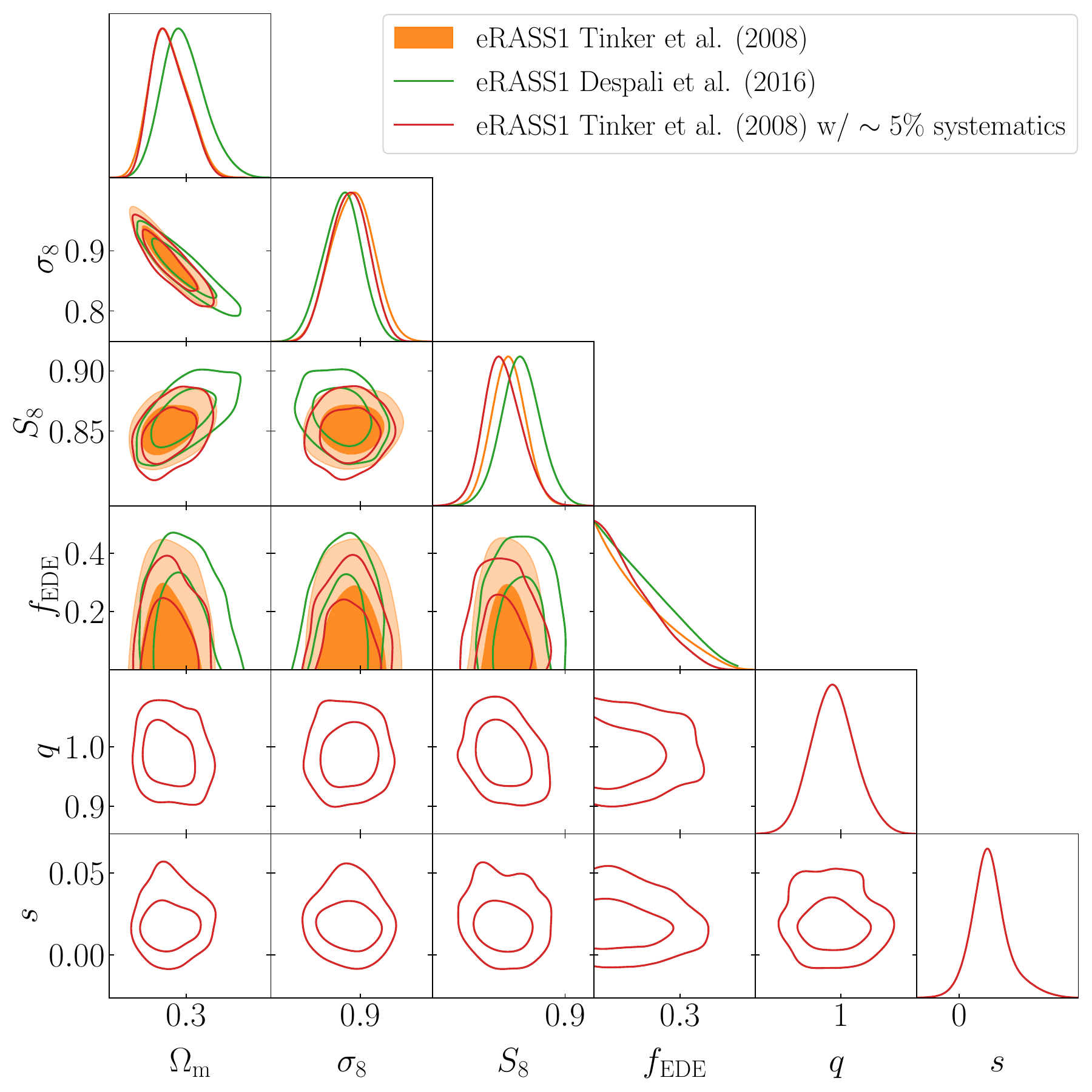}
    \caption{Robustness of our main EDE constraints against different choices of the halo mass function. We show the contour plots for the constraints on $\Om$, $\sigma_8$, $S_8$ and $f_\mathrm{EDE}$ obtained using eRASS1 $0.1<z<0.45$ for our main analysis using the \citet{Tinker2008} HMF in orange, and the \citet{Despali2016} HMF in blue. In green, we show the constraints obtained when marginalizing over additional HMF nuisance parameters, following the formalism proposed in \citet{Costanzi2019}. }
    \label{fig:FigB}
\end{figure}

We show the results of these tests in Fig.~\ref{fig:FigB}. As can be seen, neither the change to the \citet{Despali2016} HMF, nor the marginalization over additional nuisance parameters, significantly changes our constraints. The resulting posteriors for $f_\mathrm{EDE}$, $\Om$, $\sigma_8$, and $S_8$ remain fully consistent (within $<1\sigma$) with our main analysis, thus confirming that our EDE constraints are robust against current uncertainties in the halo mass function. However, as the statistical precision of cluster counts improves with future eRASS data releases, these theoretical systematics will become increasingly relevant. As discussed in Sec.~\ref{sec:discussion}, dedicated N-body simulations will be required to directly derive a $f(\sigma)$ for EDE cosmologies, or to validate at higher precisions the robustness of the modeling approach taken in this work. 

\section{EDE model sensitivity}
\label{sec:exponent_tests}

\begin{figure}[htb]
\centering
    \includegraphics[width=\linewidth]{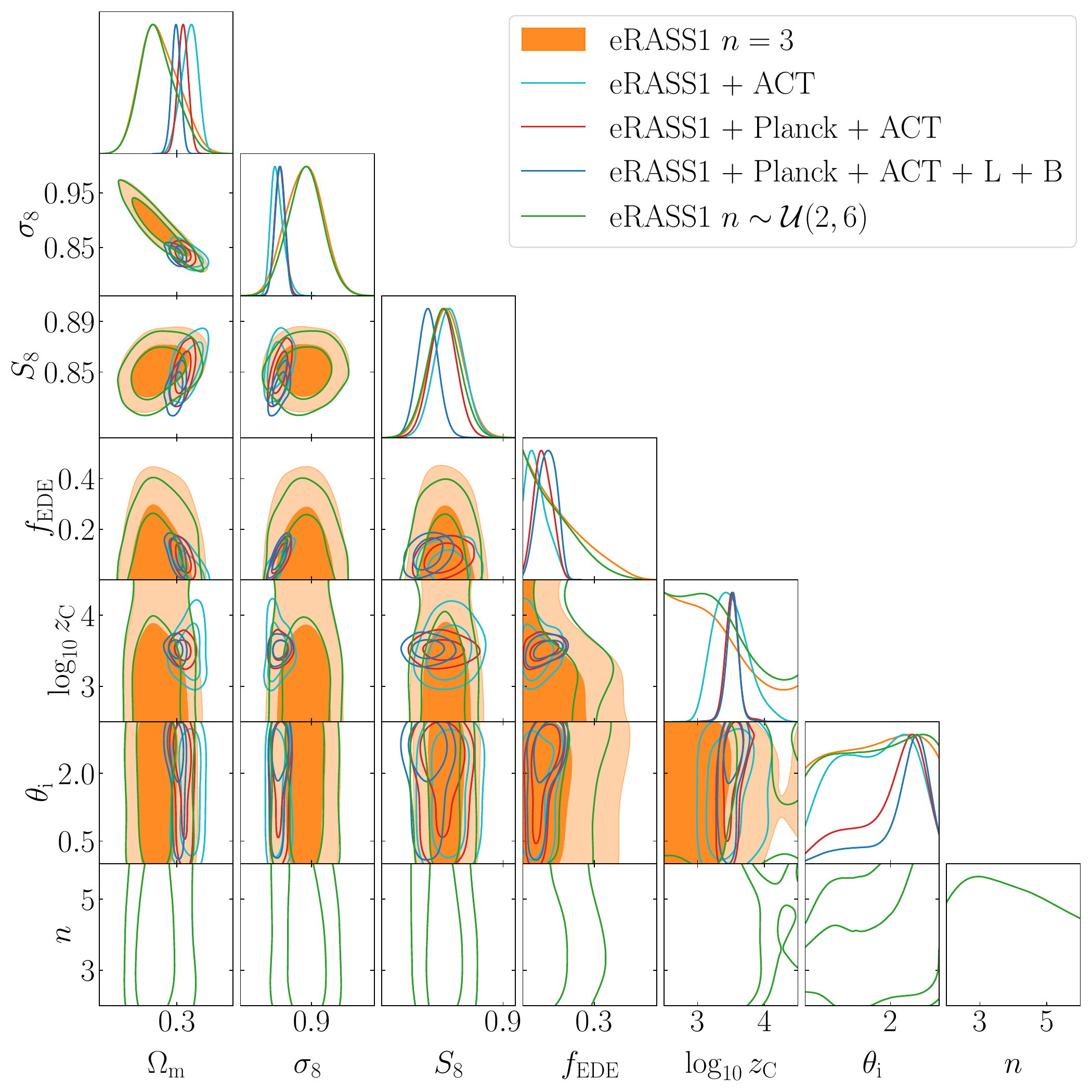}
    \caption{Extended cosmological constraints, including $\logzc$ and $\theta_i$, for the eRASS1-only and multiprobe analyses, all with $n=3$. In green, we show the constraints for the eRASS1-only analysis when leaving the axion-like potential exponent free (Eq.~\ref{eq:pot}), with a prior $n\sim\mathcal{U}(2,6)$, such as in \citet{Smith2020}. The eRASS1 constraints on structure growth cosmological parameters and on $f_\mathrm{EDE}$ are insensitive to this choice.   }
    \label{fig:FigC}
\end{figure}

We show the extended constraints obtained for the eRASS1-only and multiprobe analyses, including $\logzc$ and $\theta_i$ in Fig.~\ref{fig:FigC}. Additionally, as a sensitivity test we also show the results for our eRASS1-only analysis when leaving free the axion-like potential exponent $n$ (Eq.~\ref{eq:pot}), with a prior $n\sim\mathcal{U}(2,6)$. The figure highlights the lack of sensitivity of cluster counts to the more nuanced early-time effects of EDE, beyond establishing an upper limit on the EDE fraction via its maximal structure growth suppression. The eRASS1-only analysis provides very marginal constraints on $\logzc$, and is not sensitive to $\theta_i$, nor to the potential exponent $n$ when it is left free, with the posteriors displaying highly non-Gaussian features and being dependent on the prior choice.    

\section{Goodness-of-fit and prior volume effects}
\label{sec:gof}
\subsection{eRASS1-only analysis}
\label{subsec:PriorVolumeEffects_discussion}

To evaluate whether the inclusion of EDE in the cosmological model is able to improve the fit to cluster abundance data compared to \lcdm, we first identify the point with the maximum likelihood ($\ln\mathcal{L}_\mathrm{max}$) within the sampled parameter spaces in equivalent set-ups for both models. The difference $\Delta\chi^2 \sim -2\Delta\ln\mathcal{L}_\mathrm{max}$, with $\Delta\ln\mathcal{L}_\mathrm{max} = \ln\mathcal{L}_{\Lambda\mathrm{CDM},\,\mathrm{max}} - \ln\mathcal{L}_\mathrm{EDE,\,max}$, offers an approximate proxy for goodness of fit. However, interpreting this global eRASS1 likelihood is non-trivial; the cluster likelihood comprises multiple components, such as number counts following a Poisson distribution and log-normal scaling relations for observables, as well as the inclusion of a mixture model. Consequently, $\ln\mathcal{L}_\mathrm{max}$ does not strictly follow a standard $\chi^2$ distribution.

In agreement with the Bayesian analysis, we find $\Delta\chi^2 \sim 0.1$, consistent with no hints of a frequentist preference for EDE in our eRASS1-only analysis, despite the EDE model introducing additional degrees of freedom. While $\Delta\ln\mathcal{L}_\mathrm{max}$ would strictly be expected to be negative, this is explained by the uncertainty on this measurement---computationally expensive to robustly evaluate without significantly modifying the methodology of this work---also being expected of the same order of magnitude (as seen in similar tests performed with alternative cosmologies in \citetalias{Ghirardini2024}). This interpretation, relying on the Bayesian exploration of the parameter space, could be biased by prior volume effects (as introduced in Sec.~\ref{subsec:CMB_combination}), assuming a secondary likelihood peak exists away from the \lcdm best-fit. However, we note that in the context of EDE such a secondary peak is not strictly expected for cluster abundance data, given that its sensitivity to net structure growth suppression (showcased in Fig.~\ref{fig:Fig1}) would lead to upper limit constraints on $f_\mathrm{EDE}$, akin to its sensitivity to the summed neutrino masses $\sum m_\nu$.

To complement this global likelihood assessment, we evaluate the goodness of fit by comparing the empirical distributions of our observables against the theoretical expectations. We employ the analytical framework of the \texttt{eROCOP} pipeline to extract the one-dimensional marginalized probability densities for the X-ray count rate and photo-$z$ directly from the evaluated multi-dimensional likelihood grid, using a random subset of samples drawn from the posterior distribution to compute the mean theoretical density.

\begin{figure}[htb]
\centering
    \includegraphics[width=\linewidth]{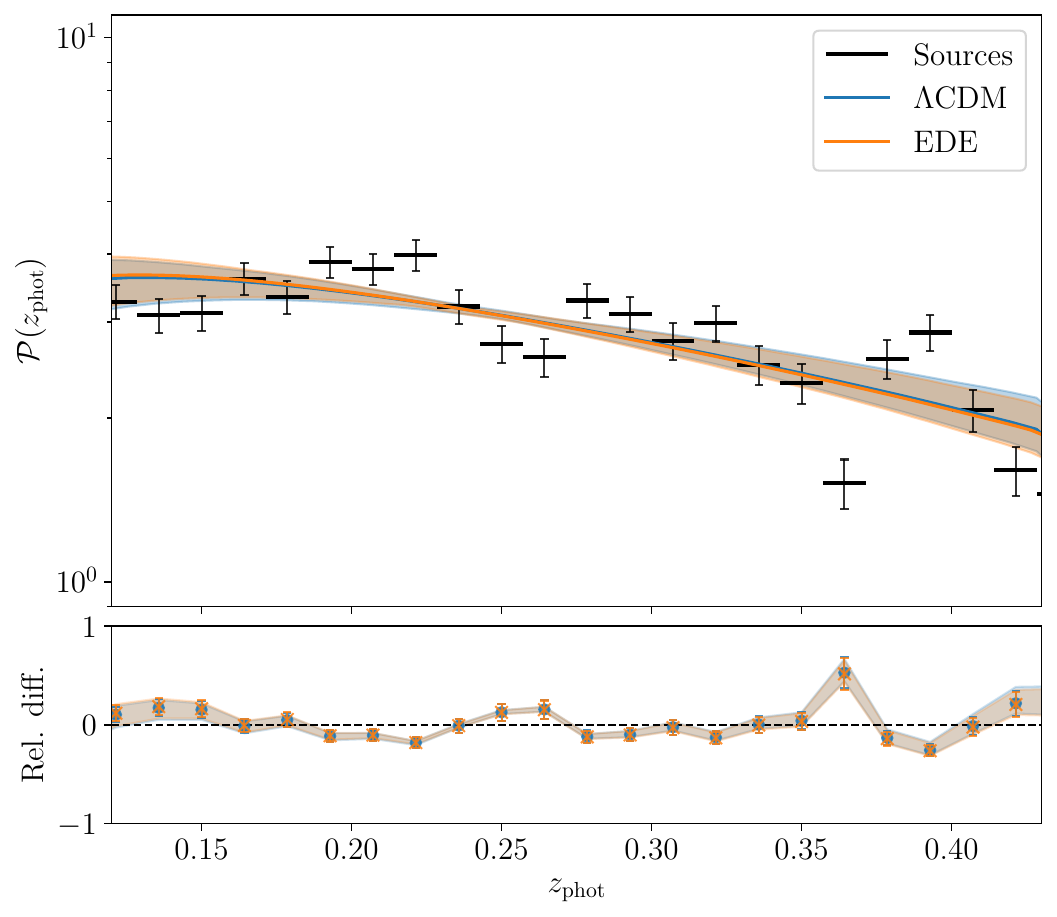}
    
    \vspace{0.1cm}
    
    \includegraphics[width=\linewidth]{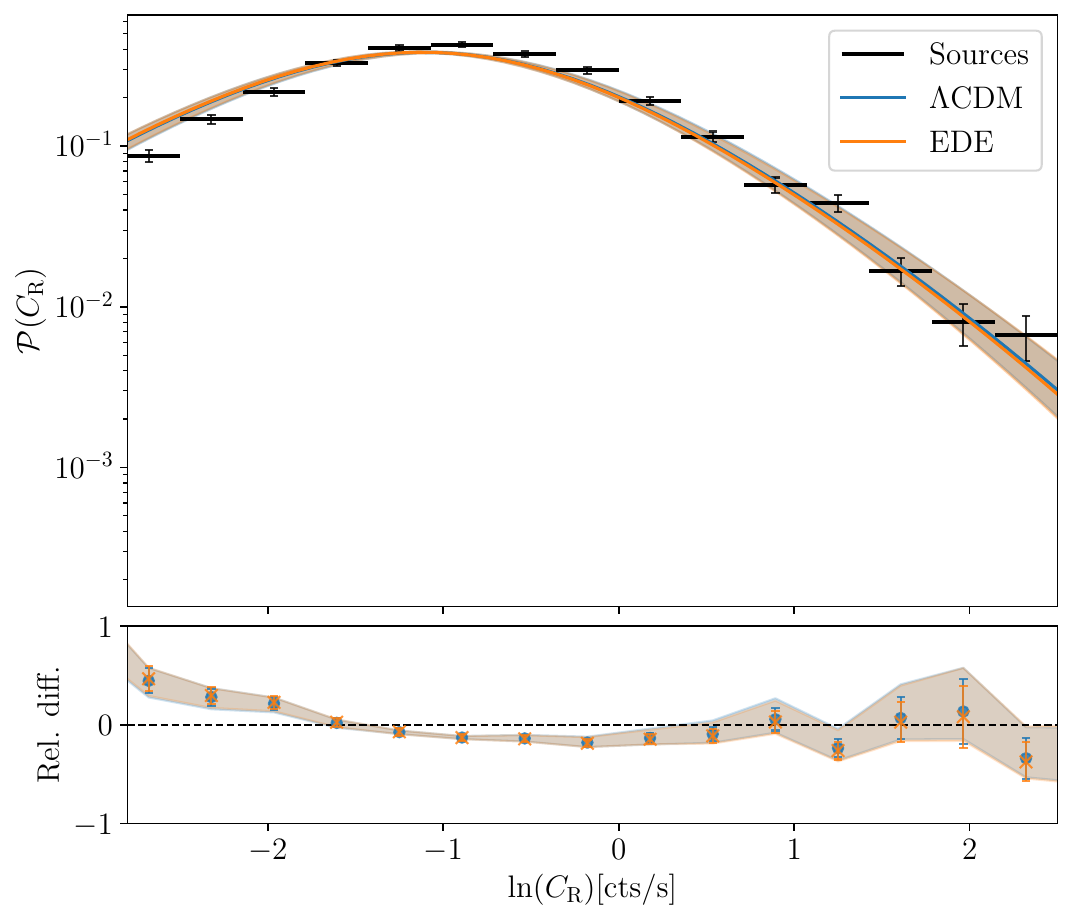}
    \caption{One-dimensional distributions of the eRASS1 cluster sample (black markers with Poisson error bars) as a function of X-ray count rate (top panel) and photometric redshift (bottom panel) for the restricted range $0.1 < z < 0.45$. The observed data are compared to the mean theoretical probability densities evaluated via the \texttt{eROCOP} pipeline for the \lcdm (blue) and EDE (orange) models. The shaded regions denote the bounds of the theoretical distributions sampled from the posterior. The lower sub-panels display the relative fractional difference between the theoretical models and the observations.}
    \label{fig:FigD1}
\end{figure}

As illustrated in Fig.~\ref{fig:FigD1}, the inclusion of EDE does not improve the fit to the observational data within our restricted redshift sample and with equivalent set-ups ($0.1 < z < 0.45$). This is expected, since our Bayesian analyses of the two models yield consistent posteriors for all shared parameters---both for the underlying cosmology and the astrophysical scaling relations (see Sec.~\ref{sec:scaling_relations})---with EDE merely suffering a marginal loss of constraining power due to its expanded parameter space. As discussed in App.~\ref{sec:exponent_tests}, these additional EDE parameters govern early-universe dynamics and provide no additional descriptive flexibility for late-time structure formation. To quantify this consistency, we compute a Poisson pseudo-$\chi^2$ statistic between the theoretical model densities and the data histograms. We note that because this analytical evaluation bypasses the forward-modeled mock catalogs used in \citetalias{Ghirardini2024}, the variance relies strictly on pure Poisson counting statistics to provide a metric for relative model comparison. Evaluated across $N_\mathrm{bins} = 22$ histogram bins for the X-ray count rates, we obtain $\chi^2 = 427.5$ for \lcdm and $\chi^2 = 456.2$ for EDE. Similarly, over $N_\mathrm{bins} = 25$ photometric redshift bins, the models are equally matched, yielding $\chi^2 = 95.2$ and $\chi^2 = 95.1$ for \lcdm and EDE, respectively.

\subsection{Multiprobe analysis}
\label{subsec:gof_multiprobe}
\begin{table*}[htb]
\caption{Model comparison metrics between EDE and \lcdm.}
\centering
\begin{tabular}{lcccc}
\hline
\hline
 &eRASS1 & +ACT & +P+ACT & +P+ACT+L+B\\
\hline
$\Delta\chi^2_{\mathrm{eff,\,eRASS1}}$ &$\phantom{-}0.1$& $-0.8$ & $-6.6$ & $-7.5$ \\
$\Delta\chi^2_{\mathrm{eff,\,ext}}$   & - & $\phantom{-}0.3$  & $-0.8$ & $-2.0$ \\
\hline
$\Delta\chi_\mathrm{eff}^2$    &$\phantom{-}0.1$& $-0.5$ & $-7.4$ & $-9.5$ \\
\hline
$\Delta\mathrm{AIC}$   & $\phantom{-}6.1$          & $\phantom{-}5.5$  & $-1.4$ & $-3.5$ \\
\hline
\hline
\end{tabular}
\tablefoot{For eRASS1, we report $\Delta\chi^2_{\mathrm{eff,\,eRASS1}}\sim-2\Delta\ln\mathcal{L}_\mathrm{max}$. For the multiprobe combinations, we identify the MAP in equivalent analyses, and separate the contributions to the posterior from the eRASS1-likelihood ($\Delta\chi^2_{\mathrm{eff,\,eRASS1}}$) and the external probes ($\Delta\chi^2_{\mathrm{eff,\,ext}}$), evaluated through a KDE at the MAP. We show the total $\Delta\chi^2_\mathrm{eff}=\Delta\chi^2_{\mathrm{eff,\,eRASS1}}+\Delta\chi^2_{\mathrm{eff,\,ext}}$, and the $\Delta\mathrm{AIC}$ (Eq.\ref{eq:AIC}). Despite the penalty for the three additional degrees of freedom in the EDE cosmology, the $\Delta\mathrm{AIC}$ values for the multiprobe analyses including \planck indicate a preference for EDE, with the majority of the goodness-of-fit improvement coming from eRASS1.}
\label{tab:gf_comparison}
\end{table*}

Because the external probes are integrated in our multiprobe analysis via a KDE of their posteriors, we cannot formally separate the Bayesian prior volume contributions from the likelihood. Consequently, unlike in the previous App.~\ref{subsec:PriorVolumeEffects_discussion}, we lack a pure maximum likelihood estimate. Instead, we identify the Maximum A Posteriori (MAP) to obtain $\ln\mathcal{L}_\mathrm{MAP}$, and compare it to that obtained in a \lcdm analysis with an equivalent set-up and use of observational probes (which \citetalias{Calabrese2025} also report for \lcdm). Considering the Gaussian calibration priors on the weak lensing mass calibration as independent dataset constraints, we analytically re-incorporate these calibration penalties to the eRASS1 likelihood at the MAP to reconstruct the total data likelihood. 

This approximation introduces an inherent Bayesian volume penalty to the EDE model, particularly in analyses where the posterior peaks near $f_\mathrm{EDE}\sim0$. Since in these regions the posterior peak can be dragged away from the true maximum likelihood by prior volume effects, this yields a strictly equal or worse fit to the data ($\ln{\mathcal{L}_\mathrm{MAP}}\leq\ln{\mathcal{L}_\mathrm{max}}$), in contrast with the \lcdm model inside which it is nested (where $\ln{\mathcal{L}_\mathrm{MAP}}\approx\ln{\mathcal{L}_\mathrm{max}}$ should hold under flat cosmological priors). 

By introducing additional EDE parameters, one would expect a decrease in $\Delta\chi_\mathrm{eff}^2\equiv-2\Delta\ln\mathcal{L}_\mathrm{MAP}$, as long as $\Delta\ln\mathcal{L}_\mathrm{MAP}\approx\Delta\ln \mathcal{L}_\mathrm{max}$. As suggested in e.g., \citet{Schoneberg2022}, to provide a fair model comparison of EDE and \lcdm we employ the Akaike Information Criterion \citep[AIC;][]{Akaike1974}, which introduces a penalty term to the additional degrees of freedom:

\begin{equation}
\label{eq:AIC}
    \Delta\mathrm{AIC}\equiv -2\Delta\ln\mathcal{L}_\mathrm{MAP}+2N_\mathrm{param}
\end{equation}

With the inclusion of the ACT DR6 posteriors, our results show that in comparison to \lcdm, the EDE model is able to provide a slightly better fit to eRASS1 data, but marginally worse for ACT. While the total $\Delta\chi_\mathrm{eff}^2$ is negative, the AIC penalization of the additional parameters indicates a preference for \lcdm from the eRASS1 and ACT probe combination.

A significant change is found in our joint analysis with CMB data from ACT and \planck. The AIC shows the EDE model provides a significantly better fit to eRASS1 data in the joint analysis, as well as modestly improving the fit to the external CMB datasets. Our full probe combination analysis yields a moderate improvement in the EDE fit for both eRASS1 and the external datasets. 

Consequently, despite the additional parameter and potential Bayesian prior volume penalizations, and anchored by the improvement for eRASS1 data, the joint analyses show a hint of preference of the EDE model over \lcdm. In this sense, the anchor provided by the mildly larger $S_8$ from eRASS1 seems to have the same effect as the $H_0$ anchor once \citetalias{Riess2022}-calibrated SN~Ia data is included in multiprobe analyses \citep[e.g.,][]{Smith2020, McDonough2024, Poulin2025}, motivating the Bayesian exploration of the $f_\mathrm{EDE}\sim0.1$ parameter space where prior volume effects are less dominant. 

The posteriors from \citetalias{Calabrese2025} we combined with include BAO and CMB lensing data jointly, and thus formally isolating their individual influence is not straightforward. We discuss their disentangled role in App.~\ref{subsec:CMBLensingBAO_Discussion}. We note that a profile likelihood analysis would provide a robust frequentist perspective as to which extent parameter space regions of $f_\mathrm{EDE}\sim0.1$ are able to fit the combined data better. Due to the computational costs of evaluating the complex eRASS1 likelihoods for such an analysis, as well as the method of integration of the external probes through their posteriors, such an approach is outside the scope of this work. Provided the hint of detection of EDE is also recovered in Bayesian analyses of upcoming eRASS data, this is left as a compelling avenue of further study.

\subsubsection{The role of CMB lensing and BAO}
\label{subsec:CMBLensingBAO_Discussion}
\begin{figure}[htb]
\centering
    \includegraphics[width=\linewidth]{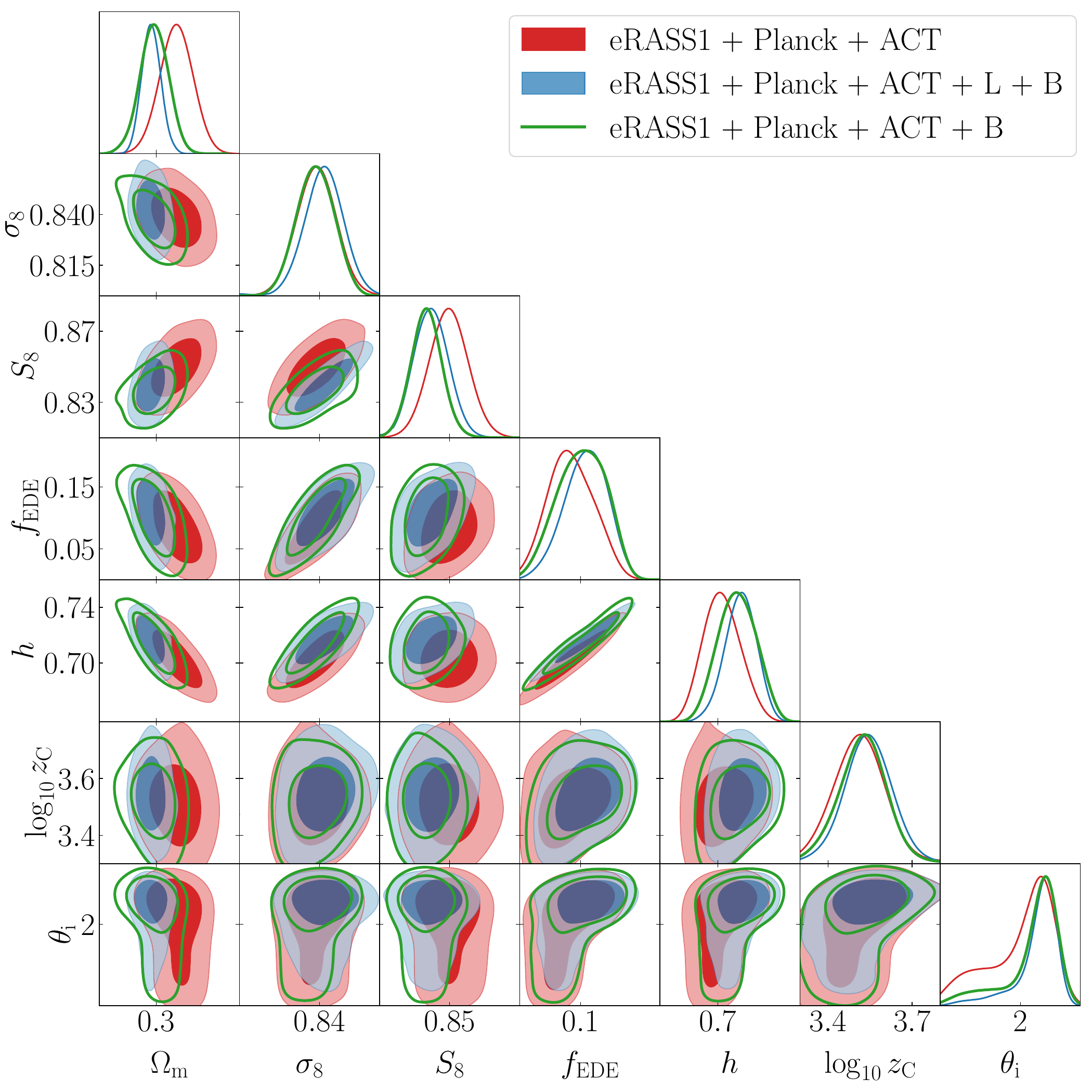}
    \caption{Extended constraints obtained with the eRASS1 analysis combined with CMB posteriors from \planck and ACT, shown in red, as well as the further combination with \planck and ACT CMB lensing and DESI BAO, shown in blue. In green, we show the constraints obtained with the combined eRASS1 and CMB analysis together with DESI BAO distance-to-sound-horizon constraints \citep{Adame2025} as priors, which are functionally equivalent to those obtained further including CMB lensing.}
    \label{fig:FigD2}
\end{figure}

We seek to distinguish the role of BAO and CMB lensing data in our full multiprobe analysis to understand which of these datasets is driving the moderate improvement in the detection significance of EDE and the goodness of fit, as well as the reduction in the residual discrepancy with late-time expansion measurements. To this end, we introduce independent priors on the distance-to-sound-horizon ratio using the latest DESI BAO measurements \citep{Adame2025} in our combined analysis with CMB data to obtain joint results for eRASS1, \planck, ACT and DESI BAO. As can be seen in Fig.~\ref{fig:FigD2}, although this dataset combination loses marginal constraining power on $\Om$, it showcases similar shifts in the marginalized posteriors to those seen in our full probe analysis (which further includes CMB lensing,  Sec.~\ref{subsec:CMB_LSS_combination}). This highlights that the $\sim1\sigma$ shifts towards lower $\Om$, $S_8$, and higher $H_0$, $f_\mathrm{EDE}$ are driven by the inclusion of DESI BAO data, in good agreement with similar shifts seen in the literature \cite{Khalife2025,Kochappan2025,Poulin2025}.

\begin{figure}[htb]
\centering
    \includegraphics[width=\linewidth]{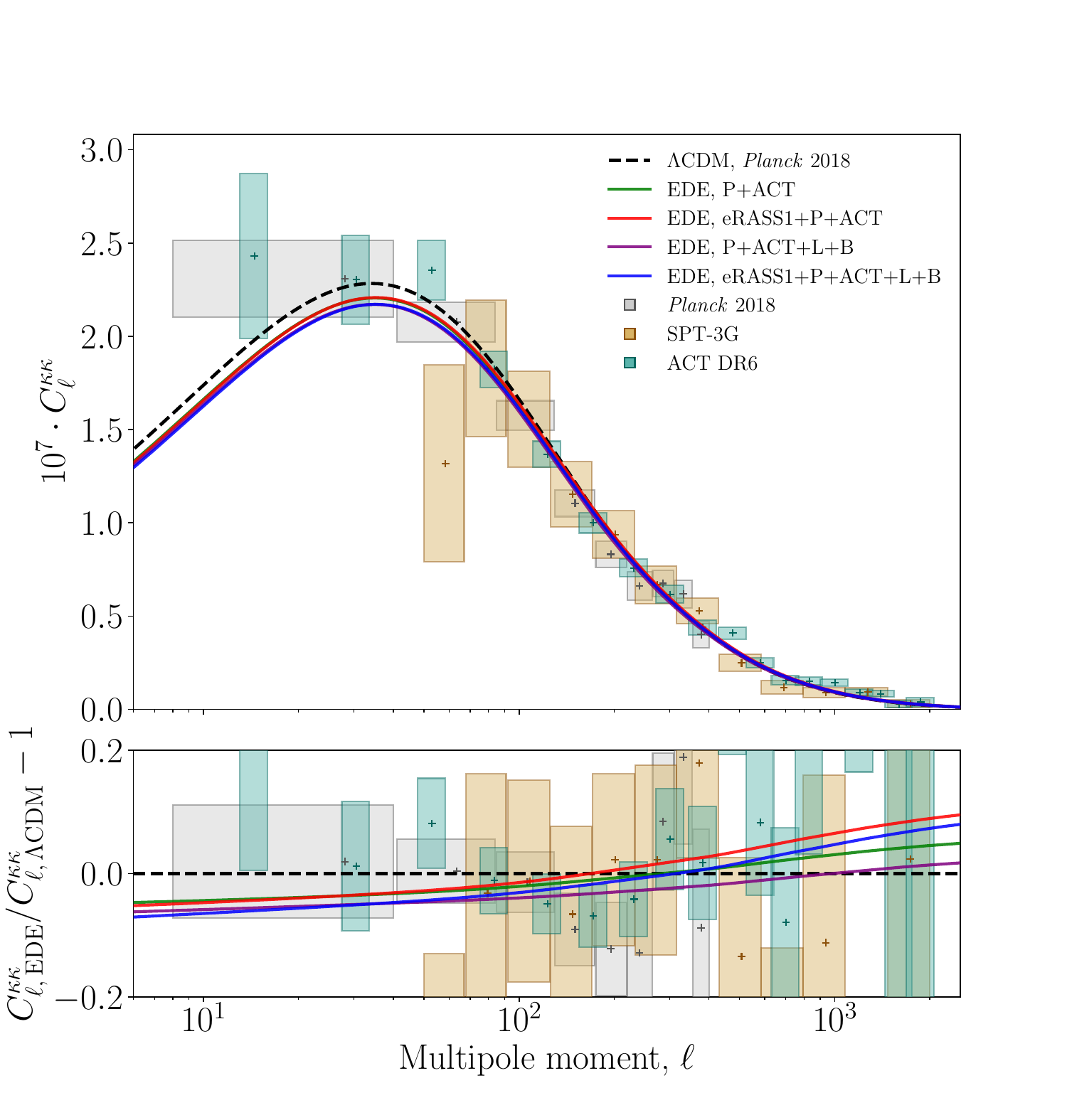}
    \caption{Comparison of the CMB lensing convergence power spectrum, $C^{\kappa\kappa}_\ell= [\ell(\ell+1)]^2 C_\ell^{\phi\phi}/4$, predicted by various constrained EDE cosmologies against observational data. The black dashed line represents the baseline \lcdm cosmology evaluated at the \citetalias{PlanckCollaboration2020} best-fit parameters. Solid colored lines denote EDE cosmologies evaluated at the posterior modes reported in \citetalias{Calabrese2025} for P+ACT and P+ACT+L+B, and our multiprobe analyses eRASS1+P+ACT and eRASS1+P+ACT+L+B. Observational bandpowers are shown as shaded boxes, covering the respective $\ell$-bins for \planck \citep{PlanckCollaboration2020b}, SPT-3G \citep{Pan2023} and ACT DR6 \citep{Qu2024, Madhavacheril2024}. The bottom panel shows the fractional residuals of the EDE model predictions relative to the \lcdm baseline. In relation to \lcdm, EDE cosmologies predict a suppression of large-scale power and an enhancement of small-scale power. The inclusion of eRASS1 cluster counts mildly exacerbates the small-scale excess, yet the theoretical shifts induced by the tested EDE cosmologies remain well within the current observational uncertainties of the lensing data.}
    \label{fig:FigD3}
\end{figure}

Thus, the question remains what the role of CMB lensing is in our multiprobe analysis, or alternatively, how the addition of eRASS1 to the multiprobe analysis impacts the fit to lensing data. To address this, we evaluate the predicted lensing spectra from the constrained EDE cosmologies against the bandpowers obtained from ACT DR6 \citep{Qu2024,Madhavacheril2024}, \planck \citep{PlanckCollaboration2020b}, and SPT-3G \citep{Pan2023} data. We show this in Fig.~\ref{fig:FigD3}, including the lensing spectra for a subset of EDE cosmologies studied in this work evaluated at the mode of our posteriors. For reference, we also include the predicted spectra from the best-fit parameters obtained in the EDE analysis of \planck and ACT, the further combination with CMB lensing measurements and DESI BAO \citepalias{Calabrese2025}, and the \lcdm predictions from the \planck best-fit parameters \citep{PlanckCollaboration2020d}. With respect to this \lcdm baseline, a general feature of the predicted CMB lensing spectrum for EDE is a suppression of power at large scales, and an excess at small scales. This is a natural consequence of the EDE-modified power spectrum (discussed in Sec.~\ref{sec:intro} and Sec.~\ref{sec:power_spectrum_impact}) featuring more power at scales $k\gtrsim k_c$ due to the shifts towards $n_s\sim1$. 

The inclusion of eRASS1 in multiprobe analyses mildly amplifies the departures from \lcdm. The CMB-fitted cosmology (\planck + ACT, $n_s\sim0.978$; $\sigma_8\sim0.831$ ) and the one jointly obtained with eRASS1 (eRASS1 + \planck + ACT, $n_s\sim0.984$; $\sigma_8\sim0.839$) showcase equivalent large-scale power suppression, but the introduction of eRASS1 predicts a more pronounced excess of power at small scales. The comparison is similar with \planck + ACT + L + B ($n_s\sim0.981$; $\sigma_8\sim0.826$) and its posterior-combined analysis with eRASS1 ($n_s\sim0.989 ;\;\sigma_8\sim0.838$), with both predicting equivalent large-scale lensing power. The joint addition of CMB lensing and BAO reduces the small-scale differences with \lcdm, yet while the former predicts only a marginal excess, the inclusion of eRASS1 amplifies these differences. Nonetheless, and while not a goodness-of-fit analysis, one can see that the relative departures induced by the addition of EDE remain within the experimental error bars of the various recent CMB lensing measurements and the predictive strength of our cosmological constraints.

\section{Scaling relation posteriors}
\begin{figure*}[htb]
\centering
    \includegraphics[width=\linewidth]{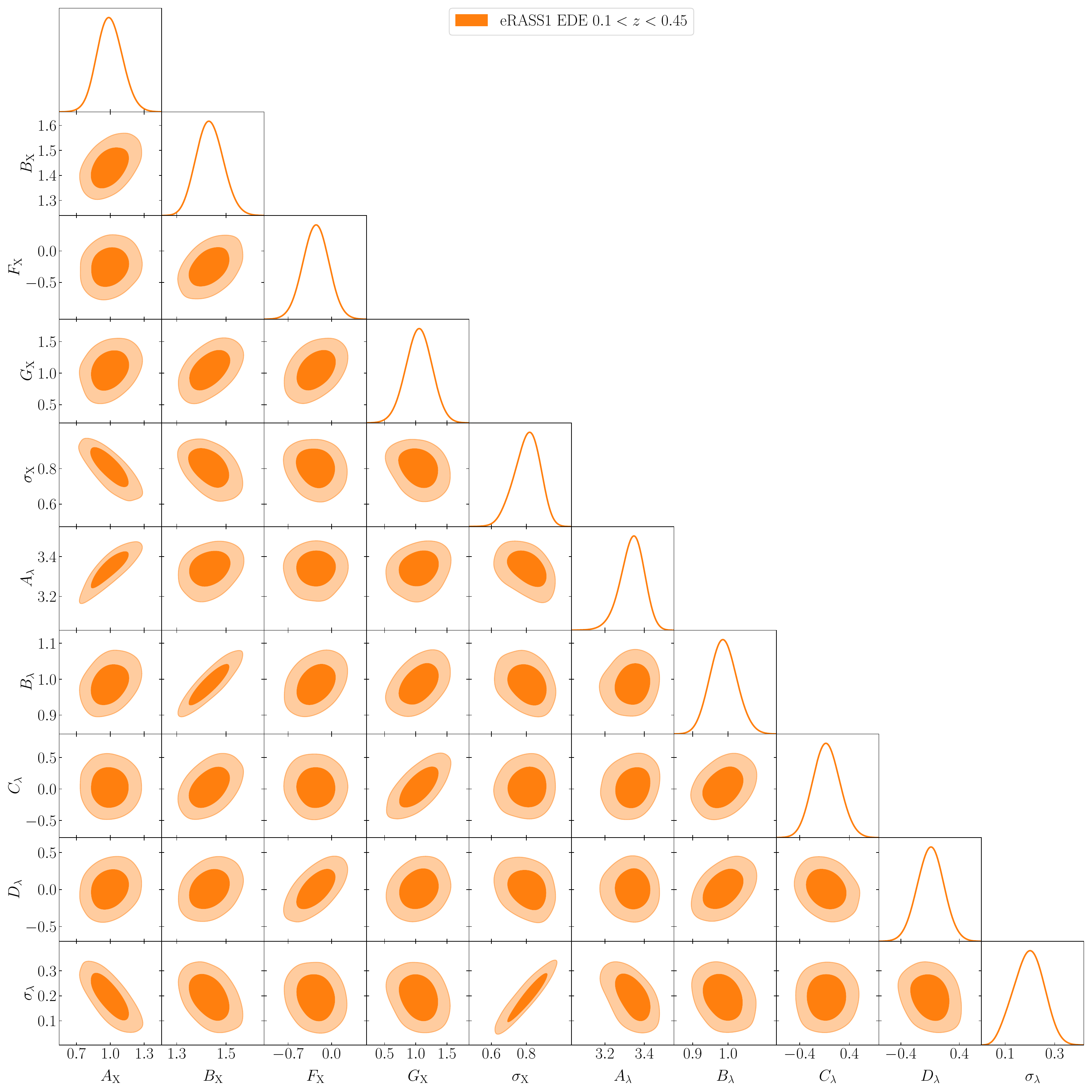}
    \caption{Posteriors on the $C_\mathrm{R}-M$ and $\lambda-M$ scaling relation parameters for the eRASS1 $0.1<z<0.45$ EDE analysis.}
    \label{fig:FigE1}
\end{figure*}

\begin{figure*}[htb]
\centering
    \includegraphics[width=\linewidth]{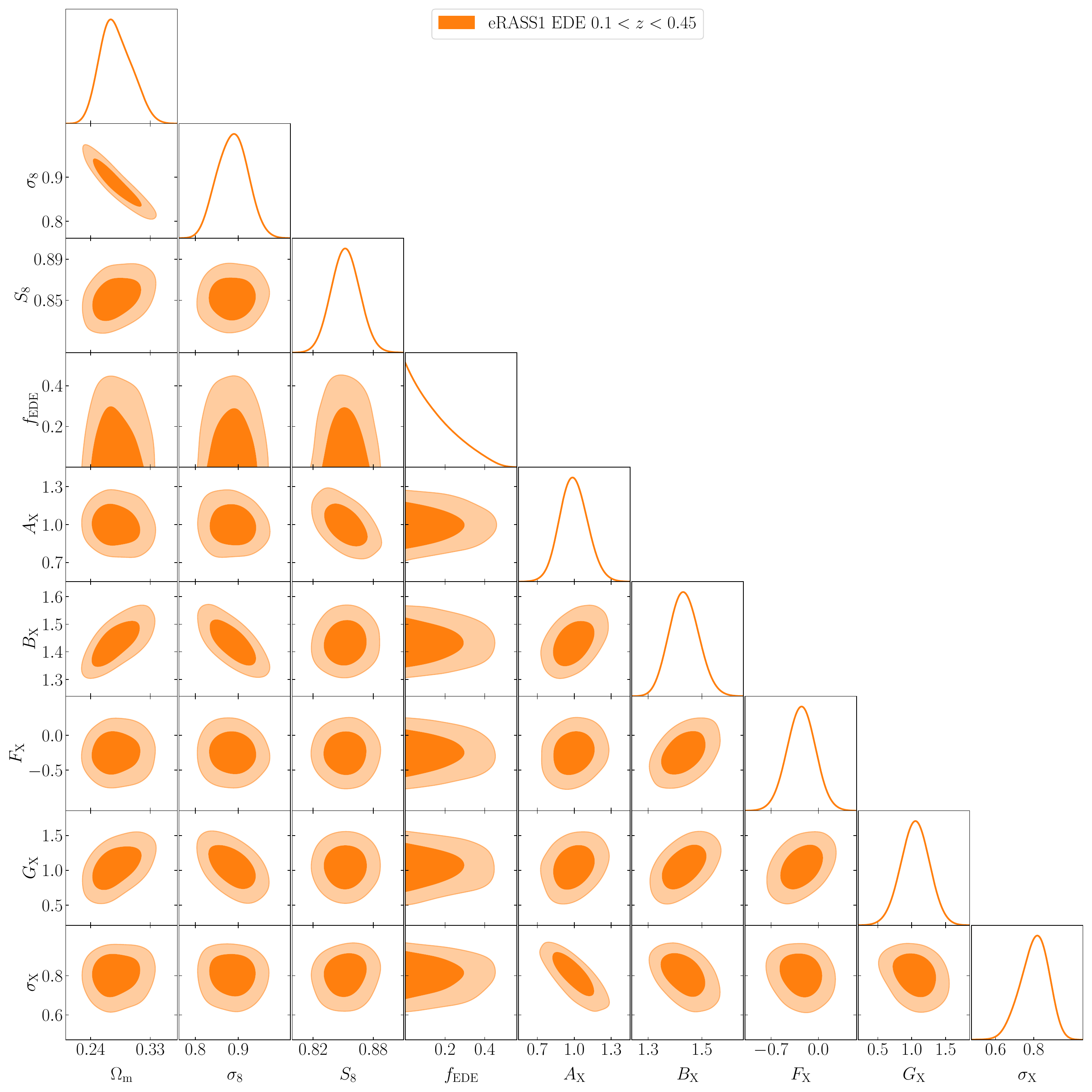}
    \caption{Degeneracies between cosmological parameters and X-ray count rate scaling relation parameters for the eRASS1 $0.1<z<0.45$ EDE analysis.}
    \label{fig:FigE2}
\end{figure*}

\begin{figure*}[htb]
\centering
    \includegraphics[width=\linewidth]{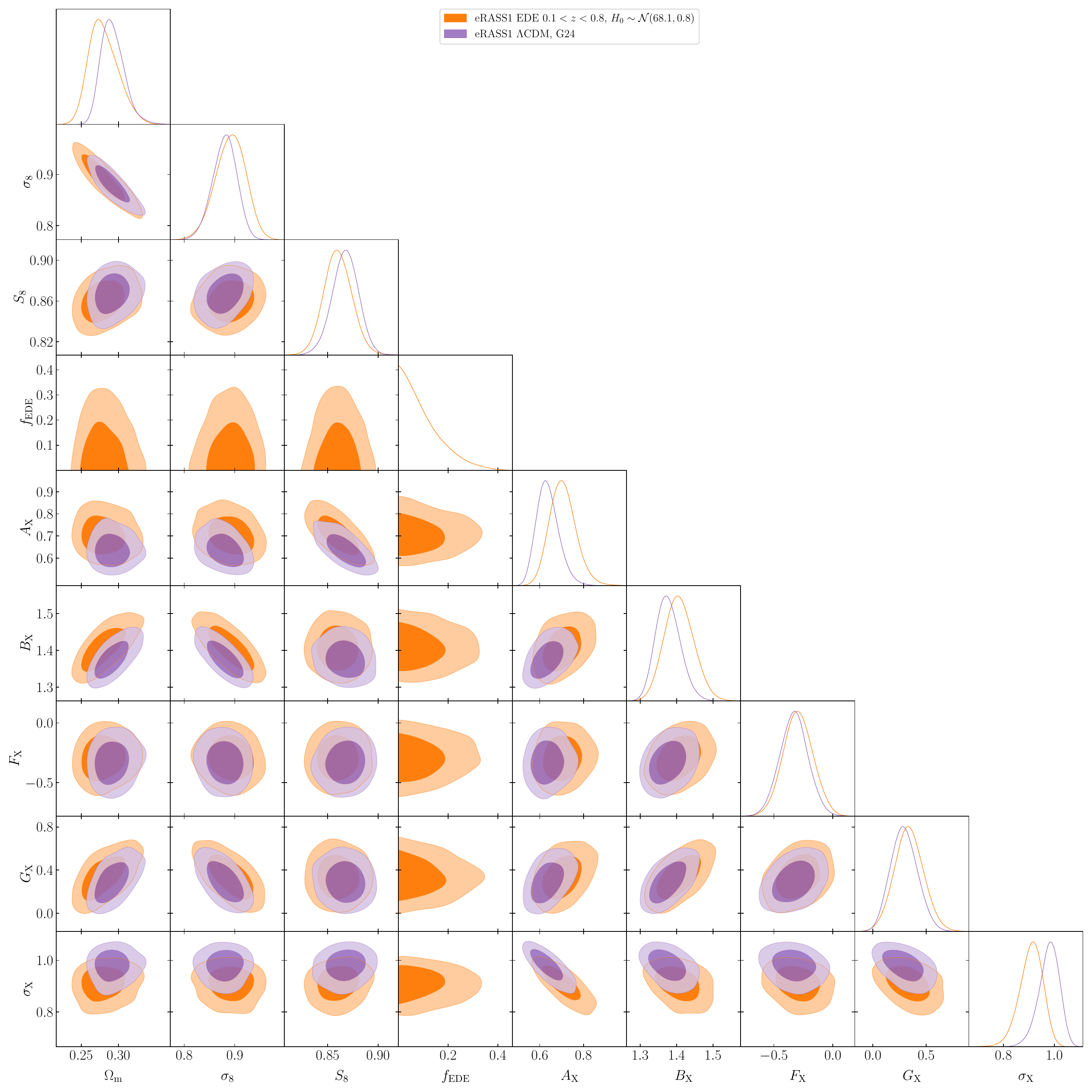}
    \caption{Posteriors on the $C_\mathrm{R}-M$ and $\lambda-M$ scaling relation parameters. In orange, we show the EDE eRASS1-only analysis with the full redshift range $0.1<z<0.8$ of the cosmology sample and using a prior on $H_\mathrm{0}$ from the \citetalias{PlanckCollaboration2020} NPIPE analysis \citep{Efstathiou2023}. In purple, we show the scaling relation parameters obtained in the \lcdm eRASS1 analysis published in \citetalias{Ghirardini2024}. }
    \label{fig:all_scaling_params}
\end{figure*}

Given that our inference pipeline jointly fits the cosmology with the scaling relation parameters required to relate the observables to the cluster mass, the introduction of EDE can lead to changes in the fitted calibration parameters in comparison to the \citetalias{Ghirardini2024} \lcdm analysis. In Fig.~\ref{fig:FigE1} we show the posteriors on the $C_\mathrm{R}-M$ and $\lambda-M$ scaling relation parameters, and in Fig.~\ref{fig:FigE2} we show the degeneracies between the X-ray count rate scaling parameters and the cosmological parameters $\Om$, $\sigma_8$, and $f_\mathrm{EDE}$. As in \citetalias{Ghirardini2024}, we see a strong correlation between X-ray and optical scaling relations. As previously discussed, the use of the reduced redshift sample impacts the scaling relation parameters and leads to lower intrinsic scatter in $\sigma_\mathrm{X}$ and $\sigma_\mathrm{\lambda}$. We also obtain larger values of the normalization of the $M-C_R$ and $M-\lambda$ scaling relations, $A_\mathrm{X}$ and $A_\mathrm{\lambda}$. We obtain a larger redshift evolution of the normalization of the $M-C_R$ scaling relation, $G_\mathrm{X}$, but parameters compatible with no redshift evolution in the $M-\lambda$ scaling relations, with the parameters $C_\mathrm{\lambda}$ and $D_\mathrm{\lambda}$ compatible with 0. Fig.~\ref{fig:all_scaling_params} shows the scaling relation posteriors for the EDE analysis of the full eRASS1 cosmology sample and with a prior on $H_\mathrm{0}$ comparable to the one used for the \lcdm analysis in \citetalias{Ghirardini2024}. This enables a comparison with a similar set-up that shows that the inclusion of EDE does not significantly modify the scaling relation parameters, where we remain compatible within $\sim1\sigma$ of those obtained in the \citetalias{Ghirardini2024} analysis.

\end{appendix}

\end{document}